\documentclass[twocolumn, a4paper, 11pt,book]{archive}

\usepackage{graphicx}
\usepackage{txfonts}
\usepackage{color}
\usepackage[dvipsnames]{xcolor}

\usepackage{txfonts}
\begin{document} 

\twocolumn[{%
 \centering
%
{\center \bf \Large Cosmic ray sputtering yield of interstellar ice mantles:}\\
\vspace*{0.05cm}
{\center \bf \Large CO and CO$_2$ ice thickness dependence.
}\\
\vspace*{0.25cm}

{\Large E. Dartois \inst{1},
        M. Chabot\inst{2},
        T. Id Barkach\inst{2},
        H. Rothard\inst{3},
        P. Boduch\inst{3}
        B. Aug\'e\inst{4}
        A.N. Agnihotri\inst{5}
       }\\
\vspace*{0.25cm}

$^1$      Institut des Sciences Mol\'eculaires d'Orsay, CNRS, Universit\'e Paris-Saclay, 
B\^at 520, Rue Andr\'e Rivi\`ere, 91405 Orsay, France\\
              \email{emmanuel.dartois@universite-paris-saclay.fr}\\
$^2$    Laboratoire de physique des deux infinis Ir\`ene Joliot-Curie, CNRS-IN2P3, Universit\'e Paris-Saclay, 91405 Orsay, France\\
$^3$   Centre de Recherche sur les Ions, les Mat\'eriaux et la Photonique, CIMAP-CIRIL-GANIL, Normandie Universit\'e, ENSICAEN, UNICAEN, CEA, CNRS, F-14000 Caen, France\\
$^4$    Institut de plan\'etologie et d'astrophysique de Grenoble, CNRS, Universit\'e Grenoble Alpes, 38000 Grenoble, France\\
$^5$    Department of Physics, Indian Institute of Technology, Hauz Khas, New Delhi, 110016, India\\           

 \vspace*{0.5cm}
{keywords: Astrochemistry, cosmic rays, molecular processes, lines and bands,  interstellar ice mantles, solid state: volatile}\\
 \vspace*{0.5cm}
{\it \large To appear in Astronomy \& Astrophysics}\\
  }]
  \section*{Abstract}
   {Cosmic-ray-induced sputtering is one of the important desorption mechanisms at work in astrophysical environments. The chemical evolution observed in high-density regions, from dense clouds to protoplanetary disks, and the release of species condensed on dust grains, is one key parameter to be taken into account in interpretations of both observations and models. 
   }
   {This study is part of an ongoing systematic experimental determination of the parameters to consider in astrophysical cosmic ray sputtering. As was already done for water ice, we investigated the sputtering yield as a function of ice mantle thickness for the two next most abundant species of ice mantles, carbon monoxide and carbon dioxide, which were exposed to several ion beams to explore the dependence with deposited energy.}
   {These ice sputtering yields are constant for thick films. It decreases rapidly for thin ice films when reaching the impinging ion sputtering desorption depth. 
   An ice mantle thickness dependence constraint can be implemented in the astrophysical modelling of the sputtering process, in particular close to the onset of ice mantle formation at low visual extinctions.}
%

\section{Introduction}
In the dense and cold zones of the interstellar medium, dust grains are covered with volatile solids (the ice mantles), which are relatively protected from ambient radiation fields outside the cloud.
However, cosmic radiation consisting of high-energy particles penetrates deeply into these clouds and induces complex chemistry through its direct interaction (radiolysis), or indirect interaction (production of a secondary vacuum ultraviolet-VUV- background radiation field by interaction with the gas, mainly molecular hydrogen) with dust grains. These cold and dense regions should in theory be considered as gas phase chemical 'dead zones' on time scales above the condensation (freeze-out) time of the whole gas phase on cold dust grains. The observation of a great diversity of species in the gas phase in these dense clouds thus implies the presence of mechanisms, not only of physico-chemical evolution, but also of desorption of the condensed species. Among these mechanisms are invoked chemical desorption \citep[e. g.][]{Yamamoto2019, Oba2018, Wakelam2017, Minissale2016, Minissale2016b, Garrod2007}, photodesorption by secondary VUV
photons or X-rays \citep{Westley1995, Oberg2009, Munozcaro2010, Oberg2011, Fayolle2011, Fayolle2013, Munozcaro2016, Cruz-Diaz2016, Cruz-Diaz2018, Fillion2014, Dupuy2017, Bertin2012, Bertin2016, Dupuy2018}, excursions in grain temperature \citep[e.g.][]{Bron2014}, and sputtering in the electronic interaction regime by cosmic radiation \citep[e.g.][and references therein]{Seperuelo2009, Dartois2013, Boduch2015, Dartois2015, Mejia2015, Rothard2017}. 
Once the efficiencies of these different processes have been quantitatively evaluated in the laboratory for a large number of systems, it is then possible to understand, through modelling, their absolute and relative impact on the chemistry of the interstellar medium.\\
The work presented in this paper is a continuation of a systematic study of the interstellar ice sputtering by cosmic radiation simulated in the laboratory. Of particular interest is the effect of the finite size of the interstellar ice mantle on the sputtering efficiency, which has been studied for H$_2$O \citep{Dartois2018}, in the context here of two of the other major ice mantle components, CO and CO$_2$.
In \S 2, the experiments of irradiation by accelerated heavy ions is described. The model of evolution of the ices under the effect of this bombardment is described in \S 3. The fitting of the obtained data for CO and CO$_2$ ices is presented in \S 4, leading to the determination of the finite sputtering depth for the considered ions. We then discuss the dependence of the sputtering depth on the stopping power and provide constraints for astrophysical models on the sputtering effective yield including finite size effects.
%
%
\begin{table*}
\caption{Experiments and results. N$_0$, $\rho,$ and {\small thickness} are the ice film initial column density, density, and thickness, respectively. $\rm \sigma_{des}$ is the radiolysis destruction cross section. Y$_s^\infty$ and N$\rm_d$ are the semi-infinite sputtering yield and sputtering depth origin of the species within the model used in this work, in a column density equivalent. l$_d$ is the depth in monolayers converted from N$\rm_d$. $\rm \sigma_s$ is the effective sputtering cross-section.
$\rm r_s/r_d$ is the ratio of the effective sputtering radius over the radiolysis destruction radius (deduced from the radiolysis destruction cross-section).}             

\begin{center}
\def\arraystretch{1.3}
\begin{tabular}{l l l l l l l l l l l}     
\hline       
%
%
{\small Species}                        &T      &N$_0$  &$\rho$$^a$     &{\small Thickness}      &$\rm \sigma_{des}$             &Y$_s^\infty$                   &N$\rm_d$                       &Depth l$_d$   &$\rm \sigma_s$ &$\rm r_s/r_d$\\
                                                &{\small K}             &{\small 10$^{16}$cm$^{-2}$}     &{\small $\rm g/cm^3$} &{\small $\rm\mu m$}     &{\small nm$^{2}$}       &{\small $\rm\times10^3$}       &{\small 10$^{16}$cm$^{-2}$}    &{\small layers}         &{\small nm$^{2}$}       &      \\ 
\hline  
\noalign{\vskip 2mm} 
\multicolumn{11}{l}{ \emph{ $H^{+}$  at 100 keV; $S_{e}$ = 46 eV / 10$^{15}$ CO$_2$ molecules/cm$^2$ \citep{Raut2013} } }\\                   
{\bf \color{RawSienna} CO$_2$}                                  &25             &50                         &1.1    &0.33                           &-                      &1.5$\pm$0.5$\times10^{-2} $       & 0.15$^{+0.76}_{-0.09}$                &1-15      &-            &-      \\
\noalign{\vskip 2mm} 
\multicolumn{11}{l}{  \emph{ $^{40}Ca^{9+}$  at 38.4 MeV ; $S_{e}$ = 1919.1 eV / 10$^{15}$ CO$_2$ molecules/cm$^2$ (this work)} } \\                   
{\bf \color{blue} CO$_2$}                               &9              &110         &1.0    &0.80           &6.3$\pm$0.6                    &15.5$\pm$3.6           & 6.7$\pm$4.6             &101$\pm$69        &26.8$^{69.6}_{12.6}$                 &2.12 \\                    
{\bf \color{red} CO$_2$}                                & 9             &58                 &1.0    &0.42           &6.8$\pm$0.5                    &13.7$\pm$8.0           &12.8$\pm$6.2           &224$\pm$108       &10.7$^{22.0}_{7.6}$           &1.26 \\
{\bf \color{green} CO$_2$}              & 9             &15.4    &1.0   &0.11           &-                                      &-                              &-                                 &-                                 &-                                     &      \\
\noalign{\vskip 2mm} 
\multicolumn{11}{l}{ \emph{ $^{58}Ni^{9+}$  at 33 MeV ; $S_{e}$ = 2487.8 eV / 10$^{15}$ CO$_2$ molecules/cm$^2$  (this work)} }\\                   
{\bf \color{blue} CO$_2$}                               & 9             &228         &1.0    &1.67           &10.7$\pm$1.9                   &48.2$\pm$17.0  &12.9$\pm$6.8           &225$\pm$119       &37.4$^{61.1}_{18.7}$          &1.87 \\               
{\bf \color{red} CO$_2$}                                & 9             &70             &1.0    &0.51           &9.2$\pm$4.7                    &37.6$\pm$23.8  &15.6$\pm$7.2           &272$\pm$126       &24.1$^{49.2}_{18.1}$          &1.47 \\
{\bf \color{green} CO$_2$}                      & 9             &38             &1.0    &0.28           &-                                      &-                              &-                                 &-                                         &-                             &-     \\
{\bf \color{Magenta} CO$_2$}                    & 9             &12.5   &1.0    &0.09           &-                                      &-                              &-                                 &-                                         &-                             &-     \\
\noalign{\vskip 2mm} 
\multicolumn{11}{l}{ \emph{ $^{58}Ni^{11+}$ at 46 MeV ; $S_{e}$ = 2601.9 eV / 10$^{15}$ C$^{18}$O$_2$ molecules/cm$^2$ \citep{Seperuelo2009} } }\\                   
{\bf \color{blue} C$^{18}$O$_2$}                        &9              &148         &1.0    &1.08           &11.7$\pm$3.1                   &78$\pm$31              &15.7$\pm$10.8          &290$\pm$200               &50.0$^{175}_{32.1}$           &2.07 \\
\noalign{\vskip 2mm} 
\noalign{\vskip 2mm} 
\multicolumn{11}{l}{ \emph{ $\rm ^{132}Xe^{21+}$ at 630 MeV ; $\rm S_{e}$ = 5680 eV / 10$^{15}$ CO$_2$ molecules/cm$^2$ \citep{Mejia2015} } }\\                   
{\bf \color{black} CO$_2$}                      &30             &6.8            &1.17   &0.042  &\multicolumn{2}{r} { $\rm{Y^{thin}_s}/{Y_{s}^\infty}\approx\frac{2.5^{+1.5}_{-0.8}10^4}{2.510^5}$ $^b$}   &65$^{+32}_{-26}$               &1023$^{+504}_{-409}$   &-                &-     \\
\noalign{\vskip 2mm} 
\noalign{\vskip 2mm} 
\multicolumn{11}{l}{  \emph{$^{40}Ca^{9+}$  at 38.4 MeV ; $S_{e}$ = 1245.2 eV / 10$^{15}$ CO molecules/cm$^2$ (this work)}} \\                   
{\bf \color{blue} CO}                           &9              &110    &0.8    &0.80           &4.4$\pm$0.4                    &32.3$\pm$10.9          &17.8$\pm$5.3           &267$\pm$79        &18.2$^{7.6}_{4.2}$            &2.03 \\                    
{\bf \color{red} CO}                            & 9             &45.5   &0.8    &0.33           &-                                      &-                              &-                                 &-                                 &-                                     &-     \\
{\bf \color{green} CO}  & 9             &19.5   &0.8    &0.14           &-                                      &-                              &-                                 &-                                 &-                                     &-     \\
{\bf \color{Magenta} CO}                        & 9             &16.7   &0.8    &0.12           &-                                      &-                              &-                                 &-                                 &-                                     &-     \\
\noalign{\vskip 2mm} 
%
\noalign{\vskip 2mm} 
\multicolumn{11}{l}{ \emph{ $^{58}Ni^{9+}$ at 33 MeV ; $S_{e}$ = 1613.2 eV / 10$^{15}$ CO molecules/cm$^2$ (this work) } }\\                   
{\bf \color{blue} CO}                   &9              &179    &0.8    &1.04           &8.9$\pm$2.1            &75$\pm$22              &15.1$\pm$9.9           &226$\pm$149            &49.9$^{138.7}_{28.4}$  &2.36   \\
{\bf \color{red} CO}                            &9              &59              &0.8    &0.34           &-                              &-                              &-                              &-                               &-                                      &-      \\
\noalign{\vskip 2mm} 
\noalign{\vskip 2mm} 
\multicolumn{11}{l}{ \emph{ $^{58}Ni^{13+}$ at 50 MeV ; $S_{e}$ = 1702.0 eV / 10$^{15}$ CO molecules/cm$^2$ \citep{Seperuelo2010} } }\\                   
{\bf \color{blue} CO}   &9              &158    &0.8    &0.92           &8.7$\pm$1.5            &83$\pm$13              &19.8$\pm$3.9           &297$\pm$58             &42.0$^{17.9}_{12.1}$   &2.20   \\
{\bf \color{red} CO}            &9              &112            &0.8    &0.65           &-                              &-                              &-                              &-                               &-                                      &-      \\
%
\hline                  
\end{tabular}
\end{center}
$^a$ The density (g/cm$^3$) considered for the ice thickness and the determination of the number of molecular layers are given for CO$_2$ for the different temperatures following \cite{Satorre2008}, and taken as 0.8 g/cm$^3$ for CO \citep[e.g.][]{Loeffler2005, Bouilloud2015}.
$^b$  Ratio of thin-to-thick yields. The yield value for infinite thickness is obtained from the fit of the other data with a power law for the considered stopping power. See text for details.
\label{table:1}
\end{table*}
%
%
\section{Experiments}
%
\begin{figure*}
\centering
\includegraphics[width=\linewidth]{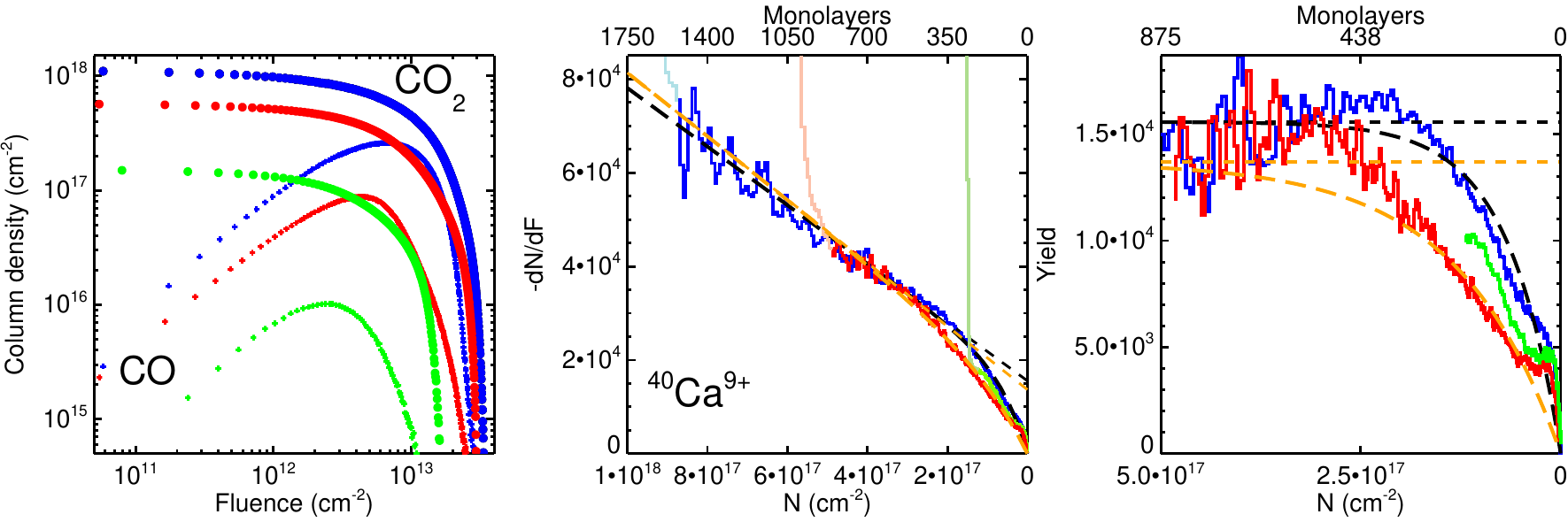}
\caption{Left panel: CO$_2$ column density evolution measured with the anti-symmetric stretching mode ($\rm\nu_3$) spectra as a function of $\rm^{40}Ca^{9+}$ ion fluence for CO$_2$ ice films deposited with different initial thicknesses (shown with different colours) measured at 9~K. The radiolytically produced CO is also shown. Middle panel: Experimentally measured differential evolution -dN/dF as a function of column density, to be compared to Equation \ref{equation_yield}. Fits of the equation to the data are shown as long-dashed lines (in black for the thickest film and orange for the intermediate thickness one) for the two thickest films, as well as fits not taking into account the finite depth of sputtering (dashed lines). No fit is attempted for the thinnest film. Right panel: Sputtering yield evolution as a function of column density; over-plotted are the infinite thickness yield (dashed lines) and adjusted exponential decay (long-dashed lines). See text for details. The experiments are summarised in Table~\ref{table:1}.}
\label{fig:Reduction_globale_CO2_18_MAY}
\end{figure*}
%
%
\begin{figure*}
\centering
\includegraphics[width=\linewidth]{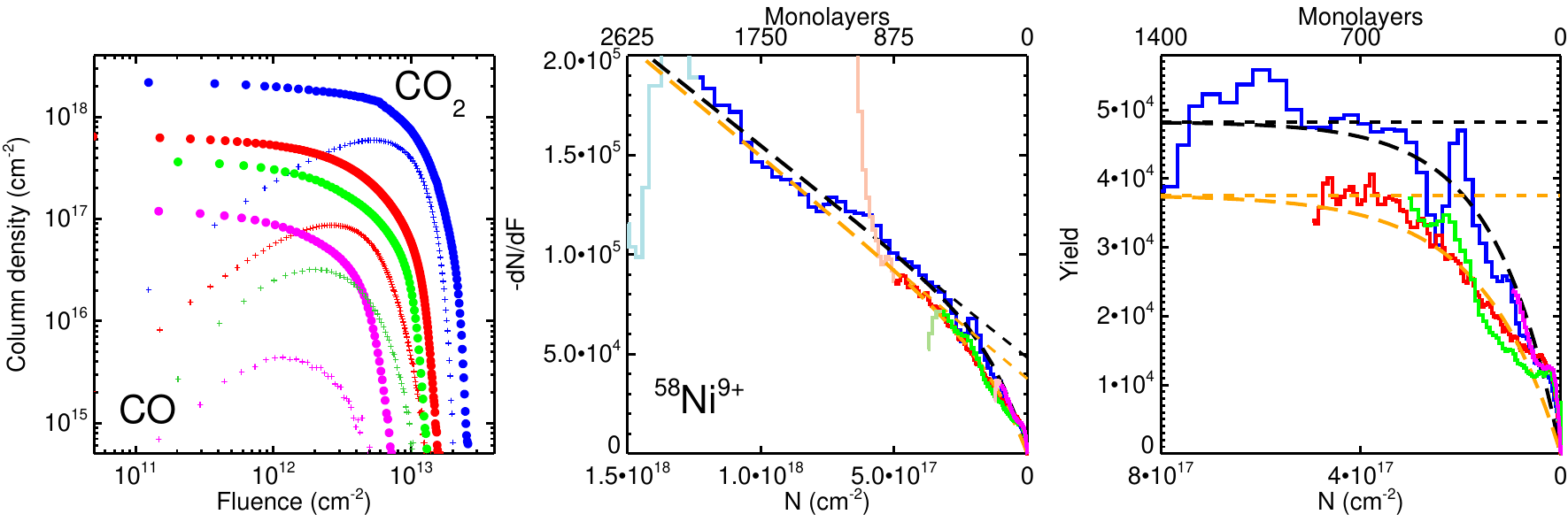}
\caption{Left panel: CO$_2$ column density evolution measured with the anti-symmetric stretching mode ($\rm\nu_3$) spectra as a function of $\rm^{58}Ni^{9+}$ ion fluence for CO$_2$ ice films deposited with different initial thicknesses (shown with different colours) measured at 9~K. The radiolytically produced CO is also shown. Middle panel: Experimentally measured differential evolution -dN/dF as a function of column density, to be compared to Equation \ref{equation_yield}. Fits of the equation to the data are shown as dashed lines for the two thickest films, as well as fits not taking into account the finite depth of sputtering (straight lines). Right panel: Sputtering yield evolution as a function of column density; over-plotted are the infinite thickness yield (dashed lines) and adjusted exponential decay (long-dashed lines). See text for details. The experiments are summarised in Table~\ref{table:1}.}
\label{fig:Reduction_globale_CO2_19_MAY}
\end{figure*}
%
\begin{figure*}
\centering
\includegraphics[width=\linewidth]{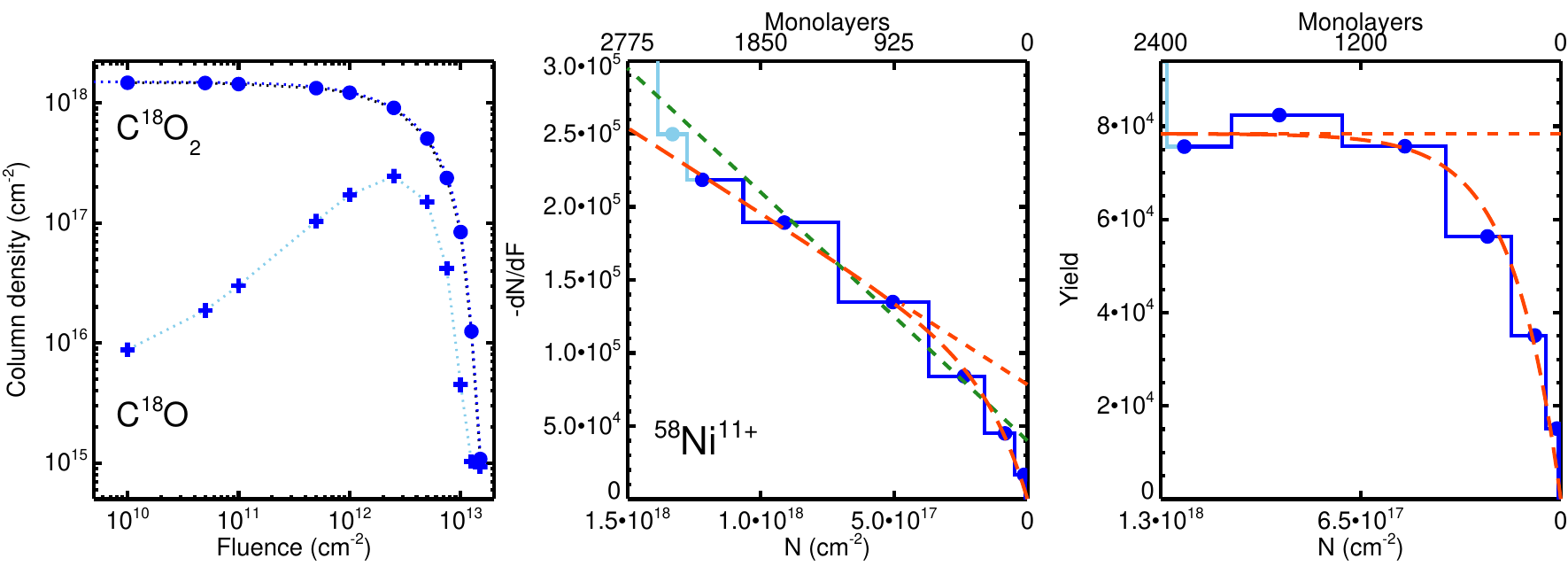}
\caption{Left panel: C$^{18}$O$_2$ ice experimental column densities for ion irradiation experiment (beam of $\rm^{58}Ni^{11+}$ at 46 MeV) discussed in \cite{Seperuelo2009}. Middle panel: $-dN/dF$ calculated from the recorded data. The blue circles represent the data used in the fit, whereas the sky blue points are discarded as they represent the phase transition of the ice observed at the early irradiation stage (low fluence). The long, dashed orange line corresponds to the best-fit models using equation \ref{equation_yield}, and the dashed orange line to what would be expected from thick film behaviour. The dotted green line represents the fit using previously determined values from \cite{Seperuelo2009}. Right panel: Sputtering yield evolution and fitted contribution as a function of column density.} 
\label{fig:Reduction_globale_CO2_18_MAY_ajustements}
\end{figure*}
%
%
\begin{figure*}
\centering
\includegraphics[width=\linewidth]{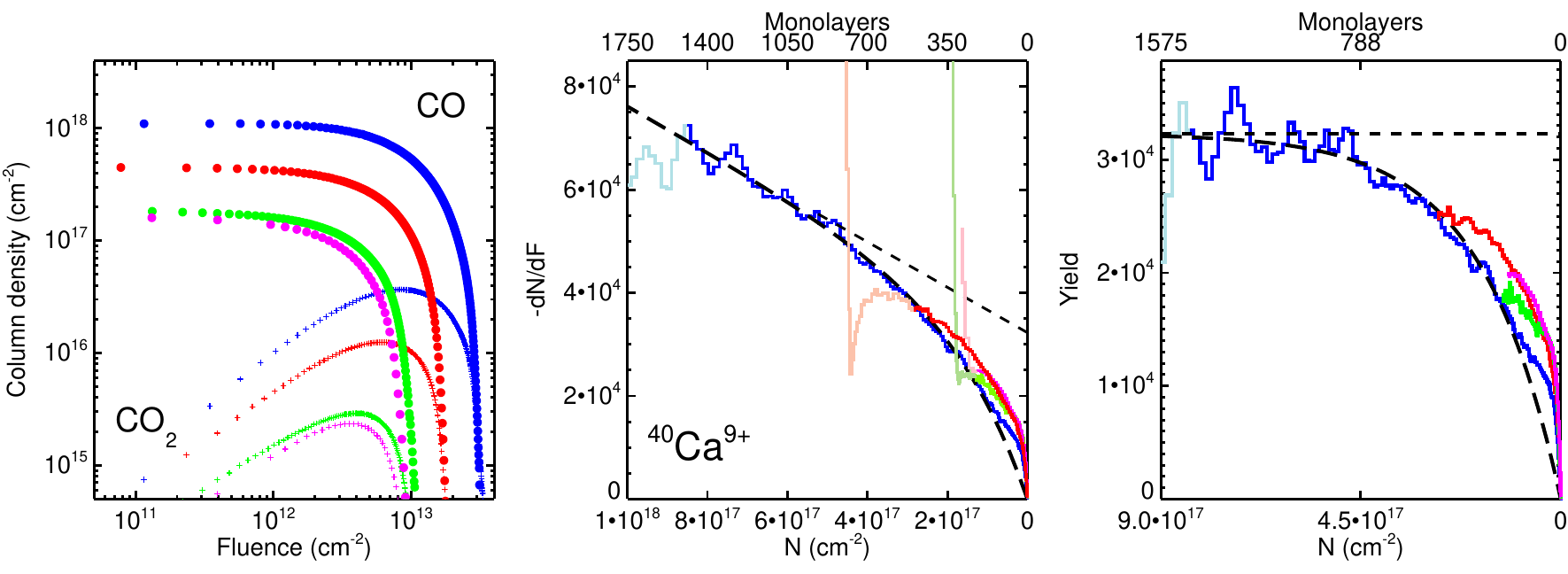}
\caption{Left panel: CO column density evolution at 9K as a function of at 33 MeV $\rm^{58}Ni^{9+}$ ion fluence for CO ice films deposited with different initial thickness (different colours). The radiolytically produced CO$_2$ is shown. Middle panel: Experimentally measured differential evolution -dN/dF as a function of column density. The fit of Equation \ref{equation_yield} to the data for the thickest film is shown (long black dashed line), as well as a fit not taking into account the finite depth of sputtering (straight dashed line). Right panel: Sputtering yield evolution and fitted contribution as a function of column density.}
\label{fig:Reduction_globale_CO_18_MAY_ajustements}
\end{figure*}
%
%
\begin{figure*}
\centering
\includegraphics[width=\linewidth]{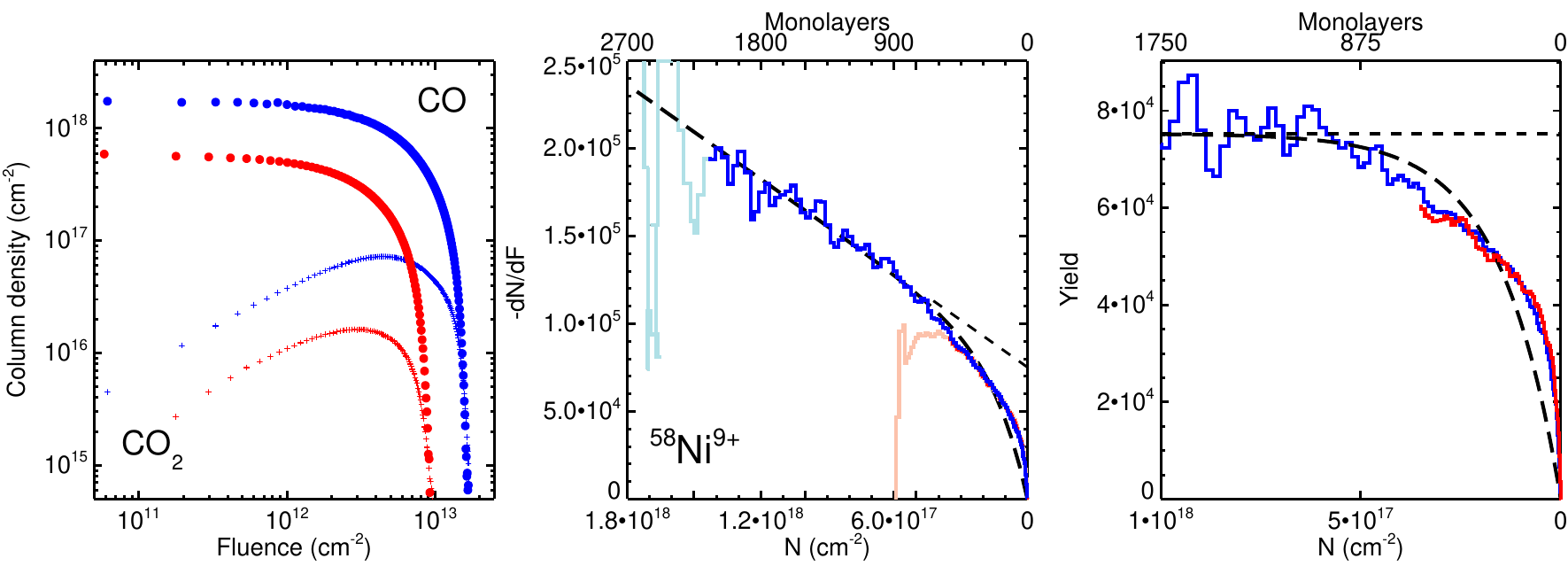}
\caption{Left panel: CO column density evolution at 9K as a function of at 33 MeV $\rm^{58}Ni^{9+}$ ion fluence for CO ice films deposited with different initial thickness (different colours). The radiolytically produced CO$_2$ is shown. Middle panel: Experimentally measured differential evolution -dN/dF as a function of column density. The fit of Equation \ref{equation_yield} to the data for the thickest film is shown (long black dashed line), as well as a fit not taking into account the finite depth of sputtering (straight dashed line ). Right panel: Sputtering yield evolution and fitted contribution as a function of column density.}
\label{fig:Reduction_globale_CO_19_MAY_ajustements}
\end{figure*}
%
%
\begin{figure*}
\centering
\includegraphics[width=\linewidth]{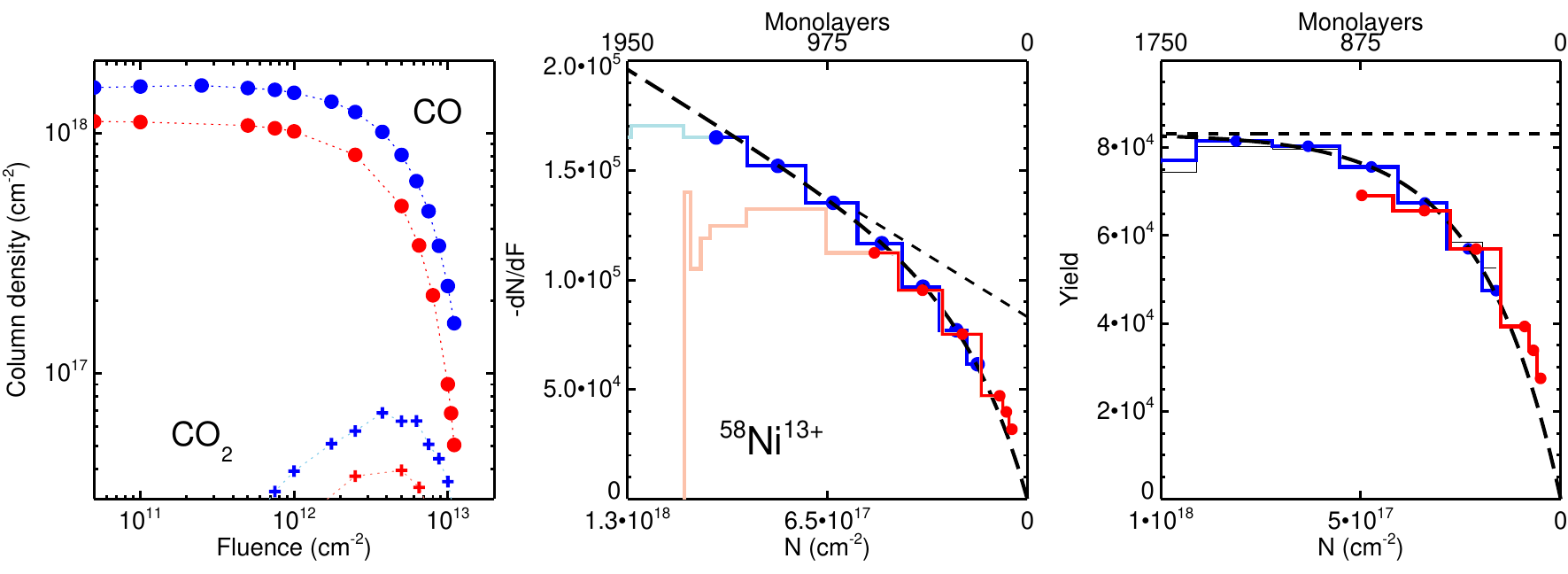}
\caption{Left: CO column density evolution at 9K as a function of fluence for 50 MeV $\rm^{58}Ni^{13+}$ ion irradiation experiments \citep[the blue one is discussed in][]{Seperuelo2009} for ice films with different initial thicknesses (different colours). Radiolytically produced CO$_2$ is shown. Middle panel: Experimentally measured differential evolution -dN/dF as a function of column density. The fit of Equation \ref{equation_yield} to the data for the thickest film is shown (long black dashed line), as well as a fit not taking into account the finite depth of sputtering (black dashed line). Right panel: Sputtering yield evolution and fitted contribution as a function of column density.} 
\label{fig:Reduction_globale_CO_08_OCT_ajustements}
\end{figure*}
%
%
Swift ion irradiation experiments were performed at the heavy-ion accelerator Grand Acc\'el\'erateur National d'Ions Lourds (GANIL; Caen, France) \footnote{Part of the equipment used in this work has been financed by the French INSU-CNRS program``Physique et Chimie du Milieu Interstellaire'' (PCMI).}.
Heavy ion projectiles were delivered on the IRRSUD beam line\footnote{http://www.ganil-spiral2.eu}. The IGLIAS (Irradiation de GLaces d'Int\'er\^et AStrophysique) facility, a vacuum chamber (10$^{-9}$ mbar under our experimental conditions) holding an infrared transmitting substrate that was cryocooled down to about 9 K, is coupled to the beam line. The ice films were produced by placing a cold window substrate in front of a deposition line. Carbon monoxide or carbon dioxide films were condensed at 9~K on the window, from the vapour phase, and kept at this temperature during the irradiations.
Details of the experimental setup are given in \cite{Auge2018}.
The ion flux, set between 10$^7$ and 10$^9$ ions/cm$^2$/s is monitored online using the current measured on the beam entrance slits defining the aperture.
The irradiation is performed at normal incidence, whereas the infrared transmittance spectra are recorded simultaneously at 12$\rm^o$ of incidence.
A sweeping device allows for uniform and homogeneous ion irradiation over the target surface.
The relation between the current at different slit apertures and the flux is calibrated before the experiments using a Faraday cup inserted in front of the sample chamber. 
The thin ice films deposited allow the ion beam to pass through the film with an almost constant energy loss per unit path length.
A Bruker FTIR spectrometer (Vertex 70v) with a spectral resolution of 1 cm$^{-1}$ was used to monitor the infrared film transmittance. The evolution of the infrared spectra was recorded as a function of the ion fluence.
The projectiles used were $\rm ^{40}Ca^{9+}$ at 38.4 MeV and $\rm ^{58}Ni^{9+}$ at 33 MeV with an electronic stopping power, calculated using the SRIM package \citep{Ziegler2010} for CO$_2$ ice of $\rm S_{e}$~$\rm=1919.1 eV / 10^{15}$ CO$_2$ molecules/cm$^2$ and $\rm S_{e}$~$\rm=2487.8 eV / 10^{15}$ CO$_2$ molecules/cm$^2$, respectively. 
We also made use of additional experiments already presented in \cite{Seperuelo2009} for C$^{18}$O$_2$ using a 46 MeV $\rm ^{58}Ni^{11+}$ projectile, a measurement on a CO$_2$ very thin film  irradiated with a 630 MeV $\rm ^{132}Xe^{21+}$ presented in \cite{Mejia2015}, and a low-energy 100~keV proton irradiation experiment on CO$_2$ from \cite{Raut2013}.
For CO ice with projectiles of $\rm ^{40}Ca^{9+}$ at 38.4 MeV, the $\rm S_{e}$~$\rm=1245.2 eV / 10^{15}$ CO molecules/cm$^2$, and for $\rm ^{58}Ni^{9+}$ at 33 MeV, the $\rm S_{e}$~$\rm=1613.2 eV / 10^{15}$ CO molecules/cm$^2$. 
 We also made use of additional experiments already presented in \cite{Seperuelo2010} for CO using a 50 MeV $\rm ^{58}Ni^{13+}$ projectile, and a low-energy 9~keV proton irradiation experiment from \cite{Schou2001}.
\section{Model}
\label{model}
As discussed previously when modelling the evolution of water ice mantles \citep{Dartois2015} {and in the modelling of a sputtering crater in the N$_2$ ice case of \cite{Dartois2020}}, the column density evolution of the ice molecules submitted to ion irradiation can be described, to first order and as a function of ion fluence (F) by a differential equation:
%
\begin{equation} 
\begin{array}{ccl}
\rm dN/dF               &\rm =  &\rm -\sigma_{des} N - Y_{s}^{\infty} \left( 1 - e^{-\frac{N}{N_d}} \right) \times f \\
\end{array}
\label{equation_yield}
.\end{equation}
%
N is the CO or CO$_2$ column density. $\rm \sigma_{des}$ is the ice effective radiolysis destruction cross-section (cm$^{2}$).
$\rm Y_{s}^{\infty}$ is the semi-infinite (thick film) sputtering contribution in the electronic regime to the evolution of the ice column density, multiplied, to first order, by the relative concentration (f) of carbon monoxide or carbon dioxide molecules with respect to the total number of molecules and radicals in the ice film.
f can be evaluated to first order by $\rm f_X=N_X/(N_{CO_2}+N_{CO})$ with X=CO$_2$ or CO, neglecting the presence of radicals, carbon suboxides, and O$_2$. 
In the case of CO$_2$, one of the main products formed by irradiation is CO, and reaches about 20\% of the ice at the top of its concentration.
When the ice film is thin (column density $\rm N\lesssim N_d$; $\rm N_d$ being the semi-infinite 'sputtering depth'), the removal of molecules by sputtering follows a direct impact model, that is, all the molecules within the sputtering area defined by a sputtering 'effective' cylinder are removed from the surface. 
The apparent sputtering yield, as a function of thickness, is modelled to first order to estimate the corresponding sputtering depth by an exponential decay, leading to the $\rm 1 - e^{(-N/N_d)}$ correcting factor applied to $\rm Y_{s}^{\infty}$.
A schematic view of such a simplified cylinder approximation is shown in Fig.1 of \cite{Dartois2018}. The sputtering cylinder is defined by a radius $\rm r_s$ (defining an effective sputtering cross section $\rm \sigma_s$) and a height d (related to the measured sputtering depth). These parameters, reported in Table.~\ref{table:1}, are calculated from the measurement of $\rm N_d$ and $\rm Y_{s}^{\infty}$. These parameters also give access to the more or less prominent elongation of the sputtering cylinder, which is species- and deposited-energy-dependent. To monitor this elongation within the cylinder approximation one can calculate the so-called aspect-ratio parameter (height-to-diameter ratio of the cylinder in the semi-infinite ice film case).
This model is a simplification. The reformation of the main species via the destruction of the daughter species is, for example, neglected as a second-order effect. The shape of the sputtering crater  also influences the exact parametrisation, as shown in \cite{Dartois2020}. An evolved model may be built when more precise data is acquired. Nevertheless, $\rm N_d$ is an efficient single parameter useful in estimating the sputtering depth in a column density equivalent.\\
The column densities of the molecules are followed experimentally in the infrared via the integral of the optical depth ($\rm \tau_{\bar{\nu}}$) of a vibrational mode, taken over the band frequency range. The band strength value (A, in cm/molecule) for a vibrational mode has to be considered.  
In the case of the $^{12}$CO$_2$ $\rm \nu_3$ mode near 2342 cm$^{-1}$, we adopted 7.6$\times$10$^{-17}$cm/molecule \citep{Bouilloud2015,Gerakines1995,ldh1986}, although some measurements tend towards a higher value \citep[$\sim$1.1$\times$10$^{-16}$cm/molecule][]{Gerakines2015}. For the $^{12}$CO $\rm \nu_1$ mode near 2342 cm$^{-1}$, we adopted 1.1$\times$10$^{-17}$cm/molecule \citep{Bouilloud2015,Gerakines1995}.
The results are anchored to these adopted values and should be modified, if another reference value is favoured. 
Fits of Equation~\ref{equation_yield} are shown in the middle panels of Figs.~\ref{fig:Reduction_globale_CO2_18_MAY}-\ref{fig:Reduction_globale_CO_08_OCT_ajustements}. Best parameters were retrieved with an amoeba method minimisation to find the minimum chi-square estimate on the model function. Only the experiments thick enough to sample the infinite thickness sputtering yield can be used. The fitted output parameters, namely $\rm \sigma_{des}$, Y$\rm_S^{\inf}$, and N$\rm_d$, are reported in Table~\ref{table:1}, with the uncertainties being estimated at two times the reduced chi-square value obtained in the minimisation.
\section{Results and discussion}
\label{results}
\subsection{Radiolysis, sputtering yield, and depth determination}
The evolution of the infrared spectra upon ion irradiation shows three stages that are much better understood when the data are plotted showing $\rm dN/dF$ as a function of the column density, rather than the column density as a function of the fluence, evolving over several decades. We clearly see in Figs.~\ref{fig:Reduction_globale_CO2_18_MAY}-\ref{fig:Reduction_globale_CO_08_OCT_ajustements} that the evolution of $\rm dN/dF$ departs from the ideal model of Equation~\ref{equation_yield}, in particular at low fluence.
At the beginning of irradiation, the ice film is restructuring with the first ions impinging the freshly deposited ice film. Therefore the molecular environment and phase is modified and/or compacted. The oscillator strength of the measured transitions in the infrared and/or the refractive index of the ice are slightly changing. As a consequence, the apparent $\rm dN/dF$ evolution is rapid. At the considered stopping powers for the ions, this is stabilised after a fluence of a few 10$^{11}$ ions/cm$^2$, and the observed behaviour of $\rm dN/dF$ better follows the expectation of the model. This early phase of the irradiation cannot be safely used to monitor the column density variations as both the molecule column density and the infrared band strength vary, leading to unpredictable changes, and they are discarded from the fits used to extract the model parameters (in the figures they are represented by light colours in the $\rm dN/dF$ plots). Including these points in the fit leads to a misestimation of the radiolysis destruction cross-section. In the second evolution stage, the film can be considered semi-infinite with respect to the sputtering and $\rm dN/dF$ evolves as a slope combining the radiolysis of the bulk and semi-infinite sputtering. In the later phase, the film becomes thin with respect to the sputtering's semi-infinite sputtering depth ($\rm N_d$) of individual ions. $\rm dN/dF$ decreases accordingly with a linear and exponentially convolved behaviour.

In each experimental session, CO and CO$_2$ films with several starting thicknesses (summarised in Table~\ref{table:1}) were irradiated. This was done to estimate the reproducibility of the results, and also to check that the behaviour of the experimental results, particularly when reaching the thin film conditions, does not depend on the initial thickness. This can be seen in the middle panels of Figs.~\ref{fig:Reduction_globale_CO2_18_MAY}-\ref{fig:Reduction_globale_CO_08_OCT_ajustements}, in which the $\rm dN/dF$ evolution is the same regardless of the initial film thickness within experimental uncertainties. It is important to state that it means that in these experiments the radiolysis (e.g. production of CO for CO$_2$ films) does not accumulate enough to significantly affect the thin film's behavioural evolution. 
\subsection{Experiments from the literature}
\subsubsection{CO$_2$} 
We reanalysed the data from \cite{Seperuelo2009} on a C$^{18}$O$_2$ ice film exposed to $\rm^{58}Ni^{11+}$ ions at 46 MeV. This is shown in Fig.\ref{fig:Reduction_globale_CO2_18_MAY_ajustements}. The right panel shows $\rm -dN/dF$ as a function of the column density, with the previous fit with the parameters from \cite{Seperuelo2009} shown via a green dashed line, and the reanalysis with our current model via orange dashed lines.
Our model includes an exponential decay at low column densities (long dashed orange lines). The dashed orange (straight) line shows
what would be expected from thick film sputtering.
As discussed above, the first points at low fluence, below about 10$^{11}$ ions/cm$^2$ were discarded from our fit, as they do not follow the theoretical expected $\rm -dN/dF$ behaviour.\\
\cite{Mejia2015} reported a strong decrease of the sputtering yield for thin CO$_2$ films exposed to $\rm^{132}Xe^{21+}$ ions at 630 MeV. 
Using the adjusted expected experimental yields for semi-infinite thick films, extrapolated to the stopping power of this experiment ($\rm S_{e}$~$=$~5680 eV / 10$^{15}$ CO$_2$ molecules/cm$^2$), a crude estimate of the sputtering depth can be made. The thin film had a column density of $\rm N_{thin}=6.8\times10^{16}cm^{-2}$, and they report a measured sputtering yield of $\rm Y_{s}^{thin}\approx 2.5^{+1.5}_{-0.8}\times10^{4}$.
The sputtering yield at the same value of electronic stopping power for a semi-infinite thick ice film can be estimated from the thick film quadratic dependence, and we estimate it to be of the order of $\rm Y_{s}^{\infty}\approx2.5\times10^{5}$. 
From $\rm Y_{s}^{thin} = Y_{s}^{\infty} {\scriptstyle \times}(1 - e^{{-N_{thin}}/{N_d}})$, one can estimate $\rm N_d\approx - N_{thin} / ln (1-Y_{s}/Y_{s}^{\infty}) \approx  6.5^{+3.2}_{-2.6}\times10^{17}cm^{-2}$.\\
To provide another point at low stopping power ($\rm S_{e}$~$=$~46 eV/10$^{15}$ CO$_2$ molecules/cm$^2$), we also include the sputtering of CO$_2$ films induced by 100 keV H$^+$ at 25K from \cite{Raut2013}. In this experiment, the authors measured a yield $\rm Y_{s}^{\infty}\approx15\pm5$. They did not measure the depth $\rm N_d$. We can fix a range
of variation by assuming upper and lower bounds limiting cases. For the lower bound, all sputtered molecules are assumed to come from the first layer. For the upper bound, they come from a depth thick as the yield. The mean value is  taken as the cube root of the yield in monolayers, that is, a homogeneous 3D sputtering volume with no preferential extension along any axis. Assuming a density of 1.1~g/cm$^3$ at 25K \citep[][]{Satorre2008} for the ice thickness and number of molecular layers, this leads to $\rm N_d\approx  1.5^{+7.6}_{0.9}\times10^{15}cm^{-2}$, providing an anchor point at low stopping power.\\

\subsubsection{CO}
To provide another point at low stopping power ($\rm S_{e}$ $=$ 14 eV/10$^{15}$ CO molecules/cm$^2$), we also included the sputtering of CO film induced by 9 keV H$^+$ at 25K from \cite{Schou2001}. In this experiment, the authors measured a yield $\rm Y_{s}^{\infty}\approx38.4$.  We again fixed its range of variation by assuming that all sputtered molecules come from the first layer for the lower bound, from a depth as thick as the yield for the upper bound, and the mean as the cube root of the yield in monolayers. Assuming a density of 0.8~g/cm$^3$ at 10K \citep[][]{Loeffler2005, Bouilloud2015} for the ice thickness and the determination of the number of molecular layers, this leads to $\rm N_d\approx  2.2^{+21}_{0.6}\times10^{15}cm^{-2}$, providing an anchor point at low stopping power.

In this reanalysis of these previous literature data, CO and CO$_2$ ice film temperatures are slightly higher than the one used in our experiments. The sputtering yield has been shown to be temperature-dependent, in particular in the case of H$_2$O ice \citep[e.g.][]{Baragiola2003}, with a higher yield when the temperature increases, $\rm Y=Y_0+Y_1 exp^{(-Ea/kT)}$), with an activation energy Ea that is related to the binding energy of the ice under consideration. For water ice it becomes significant above around 90K. We thus expect that for CO$_2$ ice, in the 9-30K range considered here, the sputtering yield should be fairly constant. In the case of CO, the measured sputtering yield might be slightly higher than expected at the same temperature as the one we used in our measurements. However, as discussed above, the error bar set by our lack of knowledge on the corresponding sputtering depth for CO covers more than one order of magnitude, and we consider it most probably includes this uncertainty.
\subsection{Relationship between sputtering depth and stopping power}
%
\begin{figure}
\centering
\includegraphics[width=\linewidth]{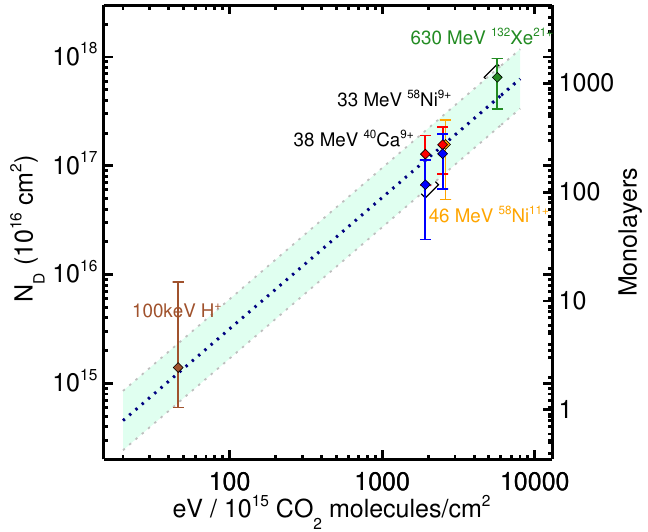}
\caption{Evolution of sputtering depth $\rm N_d$ as a function of the stopping power for CO$_2$ ice. The colour code for the points is the same as in Table~\ref{table:1}. The best fit is given by $\rm N_d= 10^{13.20\pm0.63}{\scriptstyle \times} S_{e}^{1.18\pm0.19}$. See text for details.}
\label{fig:Reduction_globale_profondeur_CO2}
\end{figure}
%
%
%
\begin{figure}
\centering
\includegraphics[width=\linewidth]{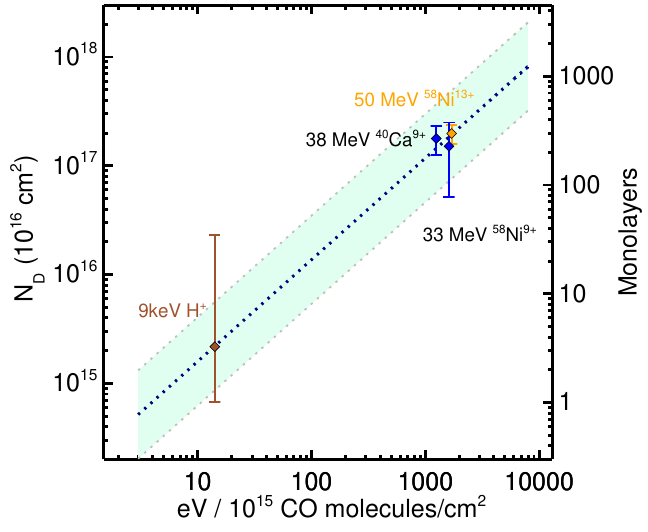}
\caption{Evolution of sputtering depth $\rm N_d$ as a function of the stopping power for CO ice. The colour code for the points is the same as in Table~\ref{table:1}. The best fit is given by $\rm N_d= 10^{14.25\pm0.46}{\scriptstyle \times} S_{e}^{0.95\pm0.17}$. See text for details.}
\label{fig:Reduction_globale_profondeur_CO}
\end{figure}
%
%
%
%
The depth of sputtering obtained from the fit with the model of Equation~\ref{equation_yield}, along with the parameters of the fit, are reported in Table~\ref{table:1}. Only the sufficiently thick films are fitted, as a sufficient number of points in the linear part of the curve is needed to retrieve the cross-section and the depth of sputtering simultaneously. The corresponding equivalent column density depth $\rm N_d$ is reported in Fig.~\ref{fig:Reduction_globale_profondeur_CO2} for CO$_2,$ and Fig.~\ref{fig:Reduction_globale_profondeur_CO} for CO as a function of the stopping power. The best fit to the data as a function of the stopping power is $\rm S_{e}^{1.18\pm0.19}$ and $\rm S_{e}^{0.95\pm0.17}$ for CO$_2$ and CO, respectively, that is, close to linear with the stopping power.
Experiments and thermal spike models of the ion-track-induced phase transformation in insulators predict a dependence of the radius r of the phase change cross-section evolving as $\rm r \sim \sqrt{S_e}$, where $\rm S_e=dE/dx$ is the deposited energy per unit path length \citep[e.g.][]{Lang2015,Toulemonde2000,Szenes1997}, with a threshold in $\rm S_e^{th}$ to be determined. 
The measured semi-infinite (thick film) sputtering yield for ices (i.e. corresponding to the total volume) generally scales as the square of the ion electronic stopping power \citep[$\rm Y_{s}^{\infty}\propto S_e^{2}$,][]{Rothard2017,Dartois2015,Mejia2015,Boduch2015}. 
In the electronic sputtering regime considered in the present experiments, and as expected from the above cited dependences, it makes sense that the sputtering depth approximatively scales linearly with the stopping power and the aspect ratio scales with the square root of the stopping power.\\

An estimate of a sputtering cross-section can be inferred from our measurements with $\rm \sigma_s\approx V/d$, where V is the volume occupied by $\rm Y_{s}^{\infty}$ molecules and d the depth of sputtering. $\rm \sigma_s\approx Y_{s}^{\infty}/l_d/ml$, where ml is the number of CO or CO$_2$ molecules/cm$^2$ in a monolayer (about $\rm 6.7\times 10^{14}$/cm$^{2}$ and $\rm 5.7\times 10^{14}$/cm$^{2}$, respectively, with the adopted ice densities).
As is shown in Table~\ref{table:1}, the sputtering radius $\rm r_s$ would therefore be about 1.26 to 2.12 times larger than the radiolysis destruction radius $\rm r_d$ in the case of the CO$_2$ ice, and 2.03 to 2.36 for CO in the considered energy range ($\sim$0.5-1~MeV/u).
The net radiolysis is the combined effect of the direct primary excitations and ionisations, the core of the energy deposition by the ion, and the so-called delta rays (energetic secondary electrons) travelling at much larger distances from the core; that is, several hundreds of nanometres at the considered energies in this work \citep[e.g.][]{Mozunder1968, Magee1980, Katz1990, Moribayashi2014, Awad2020}.
The effective radiolysis track radius that we calculate is lower than the sputtering one, which points towards a large fraction of the energy deposited in the core of the track.
The scatter on the ratio of these radii is due to the lack of more precise data. It nevertheless allows to put a rough constraint on the estimate of $\rm N_d$ in the absence of additional depth measurements, with $\rm N_d \lesssim Y_{s}^{\infty}/\sigma_d$. 
If the $\rm r_s/r_d$ ratio is high, a large amount of species come from the thermal sublimation of an ice spot less affected by radiolysis, and the fraction of ejected intact molecules is higher.
The aspect ratio corresponding to these experiments evolves between about ten and twenty
for CO$_2$ and CO, whereas for water ice at a stopping power of $\rm S_{e} \approx 3.6\times10^3 eV / 10^{15}$ H$_2$O molecules/cm$^2$, we show that it is closer to one \citep{Dartois2018}. The depth of sputtering is much larger for CO and CO$_2$ than for H$_2$O at the same energy deposition, not only because their sublimation rate is higher, but also because they do not form OH bonds. 
For complex organic molecules embedded in ice mantles dominated by a CO or CO$_2$ ice matrix, with the lack of OH bonding and the sputtering for trace species being driven by that of the matrix (in the astrophysical context), the co-desorption of complex organic molecules present in low proportions with respect to CO/CO$_2$ cannot only be more efficient, but will thus arise from deeper layers.

\subsection{Astrophysical modelling with finite size}
\begin{figure}
\centering
\includegraphics[width=\linewidth]{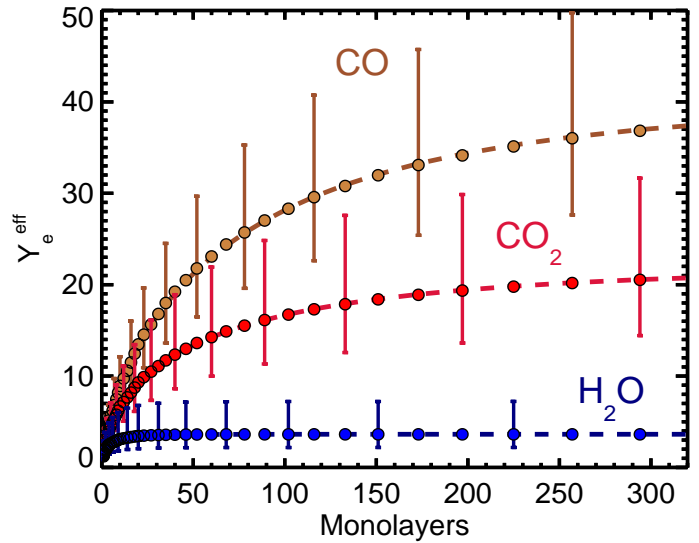}
\caption{Thickness-dependent effective sputtering yields (circles) in the electronic regime calculated for CO and CO$_2$, for $\rm \zeta_2 = 3\times10^{-17}s^{-1}$, using the thickness dependence with stopping power derived in this article. The water ice behaviour is included for comparison, calculated using the single anchor point from \cite{Dartois2018}. The functional Equation~\ref{equation_fonctionelle} fitted to the data is shown in dashed lines for each species. The typical absolute scale error bars are over-plotted.}
\label{fig:Reduction_globale_parametrisation_epaisseur}
\end{figure}

If we integrate the sputtering yield over the distribution of galactic cosmic rays (GCR), the depth dependence of the yield can be parametrised in astrophysical models to provide an effective yield. Assuming the quadratic behaviour observed for many ices \citep[e.g.][]{Rothard2017,Dartois2015,Mejia2015,Boduch2015}, 
\begin{equation} 
\rm Y^{\infty}_{s} (S_{e}) = Y_{e}^{0} S_{e}^2
,\end{equation} 
where $\rm Y_e^0$ is a pre-factor. $\rm S_{e}=(dE/dx)_e$ is the stopping power in the electronic regime. For Se in units of $\rm eV/10^{15} molecules/cm^2$, we used $\rm Y_{e}^{0} = 3.87\times10^{-2}$ for CO, $\rm 8.65\times10^{-3}$ for CO$_2$, and $\rm 5.93\times10^{-3}$ for H$_2$O.
The thickness-dependent sputtering yield in the electronic regime can be parametrised with the following prescription:
\begin{equation} 
\rm Y_{s} (N_{ice}, S_{e}) = Y_{e}^{0} S_{e}^2 {\scriptstyle \times} \left( 1 - e^{-\frac{N_{ice}}{N_d(S_{e})}} \right)
.\end{equation} 
$\rm N_{ice}$ is the ice column density, and $\rm N_d(S_{e})$ is the sputtering depth in column density equivalent, as determined in the previous section.
The effective sputtering rate by cosmic rays can be calculated by integrating over their distribution in abundance and energies:
\begin{eqnarray}\rm
\rm Y^{eff}_{e}(N_{ice}) = 2 {\scriptstyle \times} 4\pi \sum_{Z} \int_{E_{min}}^{\infty} Y_{s} [N_{ice}, S_{e}(E,Z)] \frac{dN_{CR}}{dE}(E,Z) dE 
,\end{eqnarray}
where $\rm {Y^{eff}_{e}(N_{ice})}$ is the effective sputtering rate for a given ice mantle thickness corresponding to a column density of $\rm N_{ice}$ (or equivalently a number of monolayers), $\rm \frac{dN_{CR}}{dE}(E,Z) [particles.cm^{-2}.s^{-1}.sr^{-1}/(MeV/u)]$ is the differential flux of the cosmic ray element of atomic number Z, with a cut-off in energy $\rm E_{min}$ set at 100 eV. Moving the cutoff from 10~eV to 1~keV does not change the results significantly.
The differential flux for different Z follows the GCR observed relative abundances from \citeauthor{Wang2002}~(\citeyear{Wang2002}; H, He), \citeauthor{deNolfo2006}~(\citeyear{deNolfo2006}; Li, Be), \citeauthor{George2009}~(\citeyear{George2009}; >Be), as explained in more detail in \cite{Dartois2013}.
The integration is performed up to Z=28 corresponding to Ni; a significant drop in the cosmic abundance and thus also in contribution is observed above.
The electronic stopping power $\rm S_{e}$ is calculated using the SRIM code \citep{Ziegler2010} as a function of atomic number Z and specific energy E (in MeV per nucleon).
For the differential Galactic cosmic ray flux, we adopted the functional form given by \cite{Webber1983} for primary cosmic ray spectra using the leaky box model, also described in \cite{Shen2004}:
\begin{eqnarray}\rm
\frac{dN}{dE}(E,Z) = \frac{C\;E_0^{0.3}}{(E+E_0)^3},
\label{webber}
\end{eqnarray}
where C is a normalisation constant \citep[$\rm=9.42\times10^4$,][]{Shen2004} and $\rm E_0$ a parameter influencing the low-energy component of the distribution. 
Under such parametrisation, the high-energy differential flux dependence goes asymptotically to a $-2.7$ slope. 
The ionisation rate ($\zeta_{2}$) corresponding to the same distribution can be calculated, and it gives an observable comparison with astrophysical observations in various environments. The ionisation rates  for $\rm E_0=$200, 400, and 600 Mev/u correspond to $\rm \zeta_{2} =$$3.34\times10^{-16}s^{-1}$, $5.89\times10^{-17}s^{-1}$, and $2.12\times10^{-17}s^{-1}$, respectively.
To show the dependence as a function of the number of layers, we set the value of C ($\rm E_0\approx520 Mev/u$) so that the ionisation rate corresponds to $\rm \zeta_{2} = 3\times10^{-17}s^{-1}$, a typical value for the ionisation measured in dense clouds \citep{Geballe2010, Indriolo2012, Neufeld2017, Oka2019}.\\
The best fit to the thickness-dependent effective sputtering yield, integrated over the GCR distribution corresponding to $\rm \zeta_2 = 3\times10^{-17}s^{-1}$, can be adjusted with the following functional:
\begin{eqnarray}\rm
\rm Y^{eff}_{e} (n_{layers})  \approx  \alpha{\scriptstyle \times} \left( 1 - e^{ -\left(\frac{n_{layers}}  {\beta}\right)^{\gamma}  } \right)
\label{equation_fonctionelle}
,\end{eqnarray}
with $\alpha\sim$40.1, $\beta\sim$75.8, $\gamma\sim$0.69 for CO, $\alpha\sim$21.9, $\beta\sim$56.3, $\gamma\sim$0.60 for CO$_2$, and $\alpha\sim$3.63, $\beta\sim$3.25, $\gamma\sim$0.57 for H$_2$O (plotted in brown, red, and blue in Fig.\ref{fig:Reduction_globale_parametrisation_epaisseur}, respectively).
%

The typical thickness of ice mantles estimated from astronomical observations varies from a few to several hundreds of monolayers in dense clouds
(\citeauthor{Boogert2015}~\citeyear{Boogert2015} and references therein).
In the dust particle size distribution, there might be signs of unusually large grain sizes or aggregates forming bigger particles, which is supported by the observations of scattering excess in spectral features or core shine effect \citep[e.g.][]{Jones2016, Steinacker2015, Dartois2006}.
The ice mantle growth, and thus thickness, is, to first order, independent of the grain radius, but dependent on the temperature and density.
In gas and grain astrochemical models, ice mantle thicknesses evolve with local conditions and time \citep[e.g.][]{Ruaud2016, Pauly2016}.
As the semi-infinite sputtering yield limit is reached rapidly with water ice, its effective sputtering, except at the border of clouds, should not vary significantly from the bulk, whereas for carbon dioxide and carbon monoxide, sputtering yield variations are expected when looking at the sputtering yield evolution displayed in Fig. \ref{fig:Reduction_globale_parametrisation_epaisseur}.

To calculate the sputtering rates for different ionisation rates, we varied the $\rm E_0$ parameter in Equation~\ref{webber}. It leads to a proportionality between the ionisation and sputtering rates \citep[as in, e.g.][]{Dartois2015, Faure2019}. By doing so, one assumes that the energy spectra of all CR species are identical.  For a CR spectrum with low energy particles impacting thick clouds containing the ice mantles, the stopping powers (different for each species) are differentially shaping the spectra during the propagation within the cloud \citep[see Fig. 3 in][]{Chabot2016}. To derive an estimate of the expected variations with respect to the simple proportionality between sputtering and ionisation rates, we propagated different initial CR spectra through clouds, as explained in \cite{Chabot2016}. We then calculated the ionisation rates and corresponding sputtering rates. In the worst case, the proportionality approximation  overestimates the complete calculation by a factor 2. It occurs in the very dense parts of clouds (Av>20) for initial CR spectra giving rise to high ionisation rates in the (unattenuated) external part of the cloud ($\rm \zeta_{2}>5\times10^{-16}s^{-1}$). It is worth mentioning that magnetic fields may also affect such a proportionality between ionisation and sputtering rates. Indeed, proton and heavy particles do not have the same magnetic rigidity nor Larmor radius.

\section{Conclusions}
We measured the swift, heavy ion-induced CO and CO$_2$ ice sputtering yield at $\sim$10~K, and its dependence on the ice thickness.
These measurements allow us to constrain the sputtering depth probed by the incident ion.
Within the context of an 'effective' sputtering cylindrical shape to describe the sputtered molecule volume, the aspect ratio (height-to-diameter ratio of the cylinder in the semi-infinite ice film case) is higher than one, for the ion stopping powers considered in this study. 
The ejected molecules are arising from deeper layers than would be the case for a pure water ice mantle at the same deposited energies.
The measured depth of desorption $\rm N_d$ scales with the ion electronic stopping power as $\propto$S$_e$$^{1.18\pm0.19}$ (S$_e$, deposited energy per unit path length) and $\propto$S$_e$$^{0.95\pm0.17}$, for CO$_2$ and CO, respectively. We thus experimentally measured a behaviour in agreement with what is expected from studies with swift heavy ions on insulators, as the phase transformations show a dependence of the radius r of the cross-sections evolving as $\rm r \sim \sqrt{S_e}$, and the ice's total sputtering yields are generally proportional to the square of S$_e$. 

Following the trend measured, the depth of desorption evolves almost linearly with S$_e,$ and the aspect ratio dependence will thus scale as S$_e^\alpha$, with $\alpha$$\sim$0.5. 
Astrophysical models should take into account this ice mantle's thickness dependence constraints, in particular at the interface regions and onset of ice formation, that is, when ice is close to the measured extinction threshold. We provide a parametrisation based on our experiments, which can be used to describe the sputtering yield taking into account the finite nature of interstellar grains, ready for implementation in astrochemical models.
\begin{acknowledgements}
This work was supported by the Programme National 'Physique et Chimie du Milieu Interstellaire' (PCMI) of CNRS/INSU with INC/INP co-funded by CEA and CNES, and benefited from the facility developed during the ANR IGLIAS project, grant ANR-13-BS05- 0004, of the French Agence Nationale de la Recherche. Experiments performed at GANIL. We thank T. Madi, T. Been, J.-M. Ramillon, F. Ropars and P. Voivenel for their invaluable technical assistance. 
A.N. Agnihotri acknowledges funding from INSERM-INCA (Grant BIORAD) and R\'egion Normandie fonds Europ\'een de d\'eveloppement r\'egional-FEDER Programmation 2014-2020.
%
\end{acknowledgements}


%
\begin{appendix}
\section{Infrared spectra from the experiments}
The baseline-corrected infrared optical depth spectra in the CO$_2$ $\nu_3$ antisymmetric stretching mode and CO $\nu_1$ stretching mode region for the experiments reported in this work are displayed here for the different initial films' thicknesses considered.
%
%
\begin{figure}
\centering
\includegraphics[width=\linewidth]{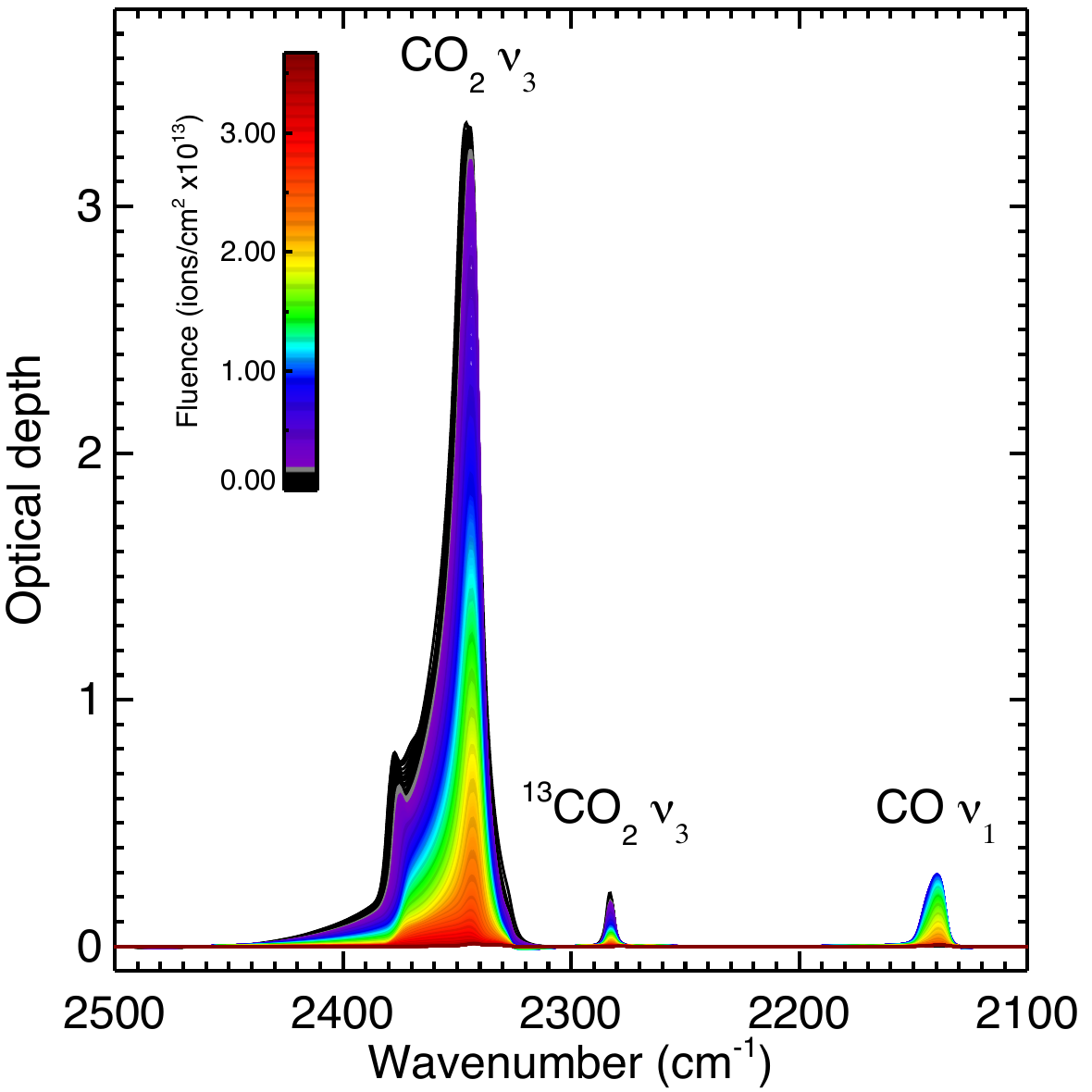}
\caption{Infrared spectra for first CO$_2$ ice experiment with $\rm^{40}Ca^{9+}$  at 38.4 MeV ions. The inserted colour code gives the corresponding irradiation fluence.}
\label{fig:180502_W2_CO2_40_Ca_9plus_0_96_MeV_u}
\end{figure}
%
%
\begin{figure}
\centering
\includegraphics[width=\linewidth]{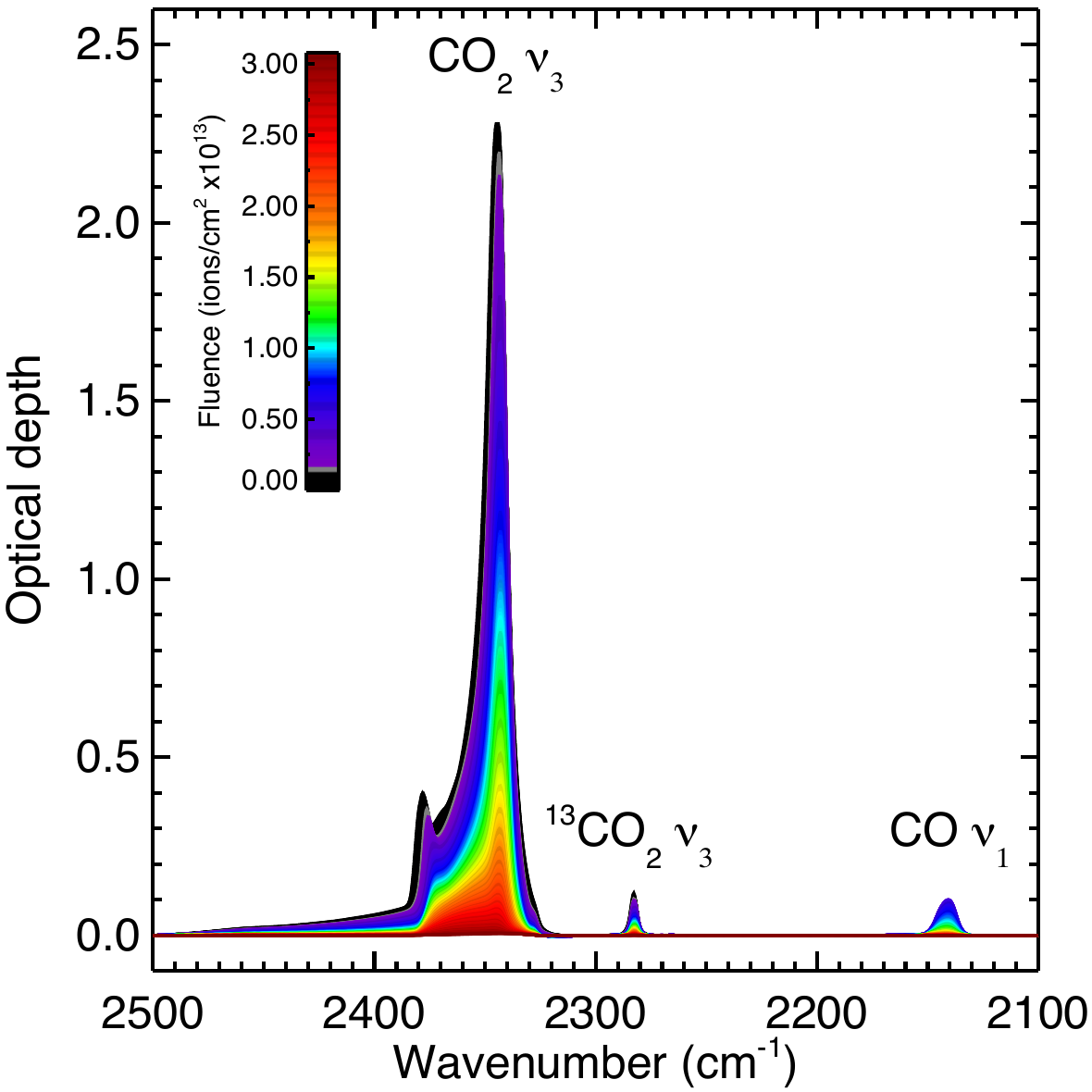}
\caption{Infrared spectra for second CO$_2$ ice experiment with $\rm^{40}Ca^{9+}$  at 38.4 MeV ions.}
\label{fig:180503_W2_CO2_40_Ca_9plus_0_96_MeV_u}
\end{figure}
%
%
\begin{figure}
\centering
\includegraphics[width=\linewidth]{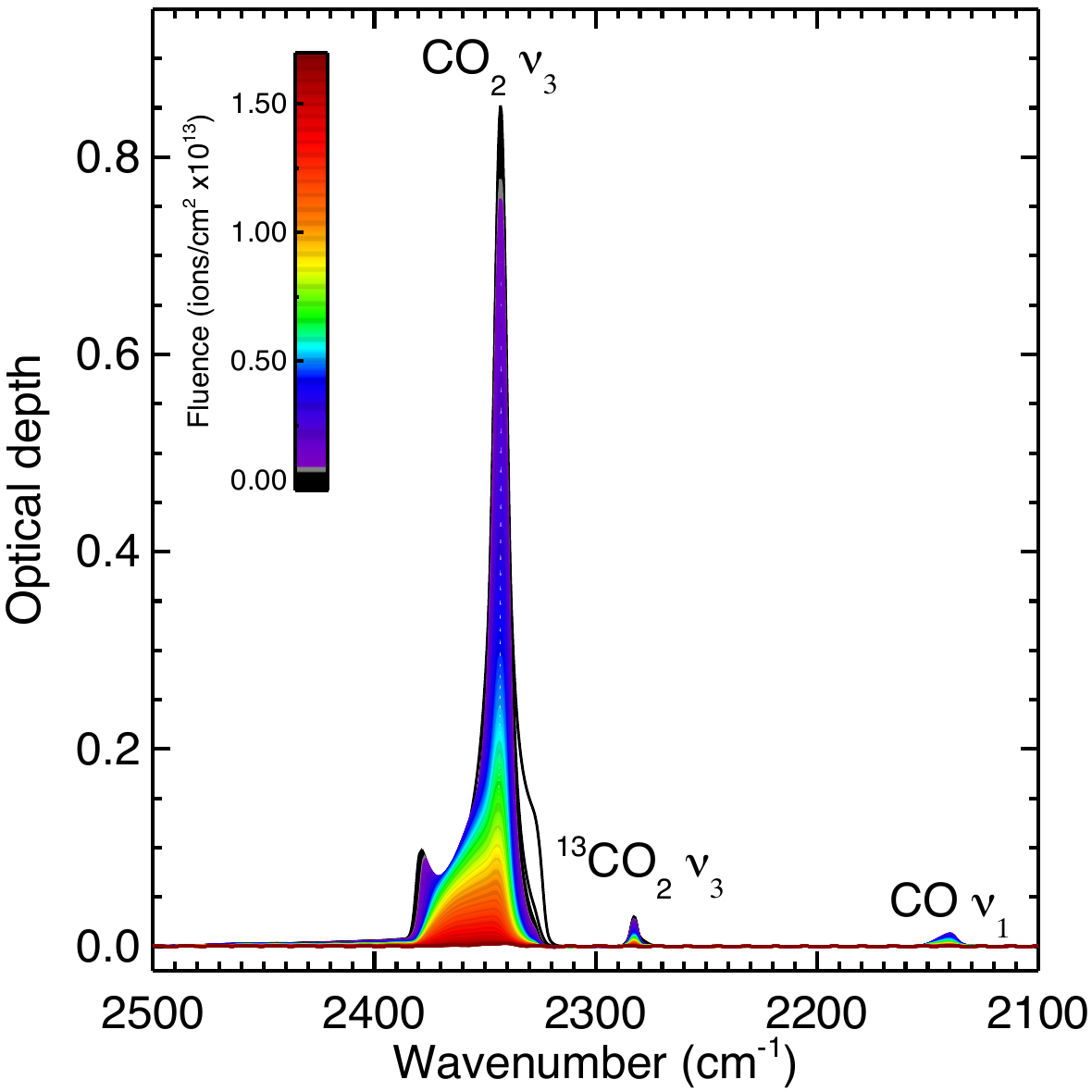}
\caption{Infrared spectra for third CO$_2$ ice experiment with $\rm^{40}Ca^{9+}$  at 38.4 MeV ions.}
\label{fig:180502_W2_BIS_CO2_40_Ca_9plus_0_96_MeV_u}
\end{figure}
%
%
%
\begin{figure}
\centering
\includegraphics[width=\linewidth]{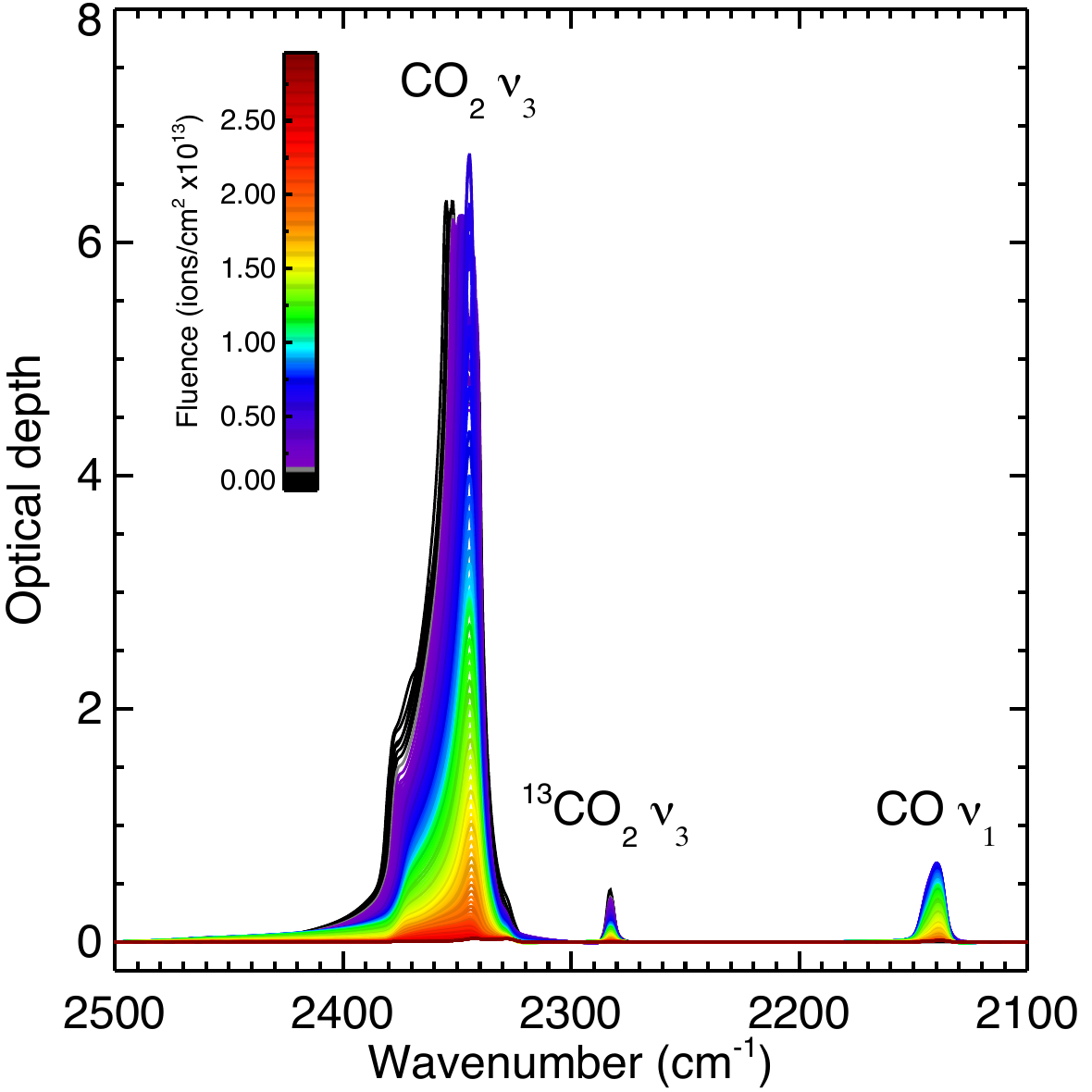}
\caption{Infrared spectra for first CO$_2$ ice experiment with $\rm^{58}Ni^{9+}$ at 33 MeV ions. We note that the spectra are saturated at the beginning of the experiment. These spectra were discarded from the analysis.}
\label{fig:19_05_02_A_W1_CO2_58_Ni_9plus_0_57_MeV_u}
\end{figure}
%
%
\begin{figure}
\centering
\includegraphics[width=\linewidth]{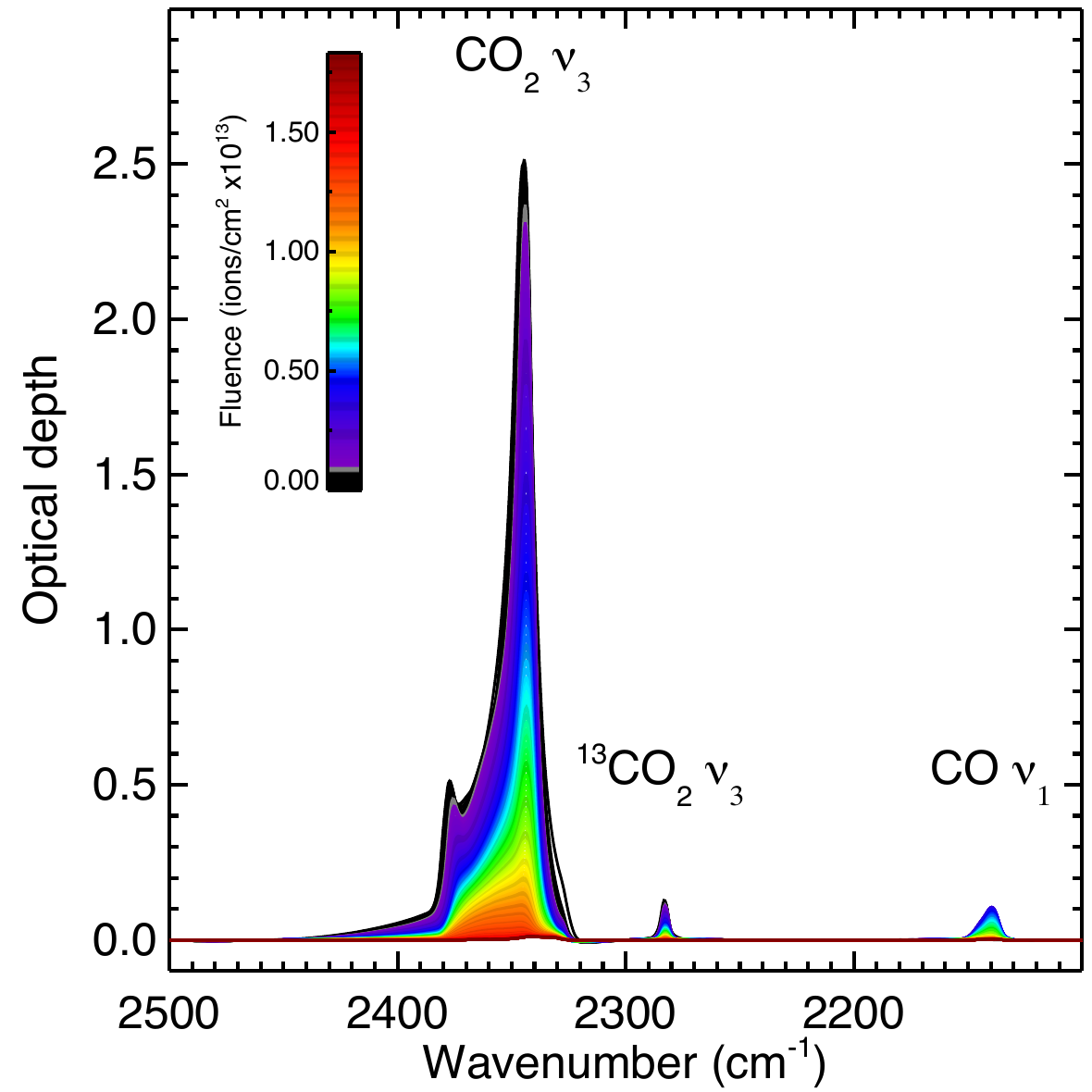}
\caption{Infrared spectra for second CO$_2$ ice experiment with $\rm^{58}Ni^{9+}$ at 33 MeV ions.}
\label{fig:19_05_04_C_W1_TER_CO2_58_Ni_9plus_0_57_MeV_u}
\end{figure}
%
%
\begin{figure}
\centering
\includegraphics[width=\linewidth]{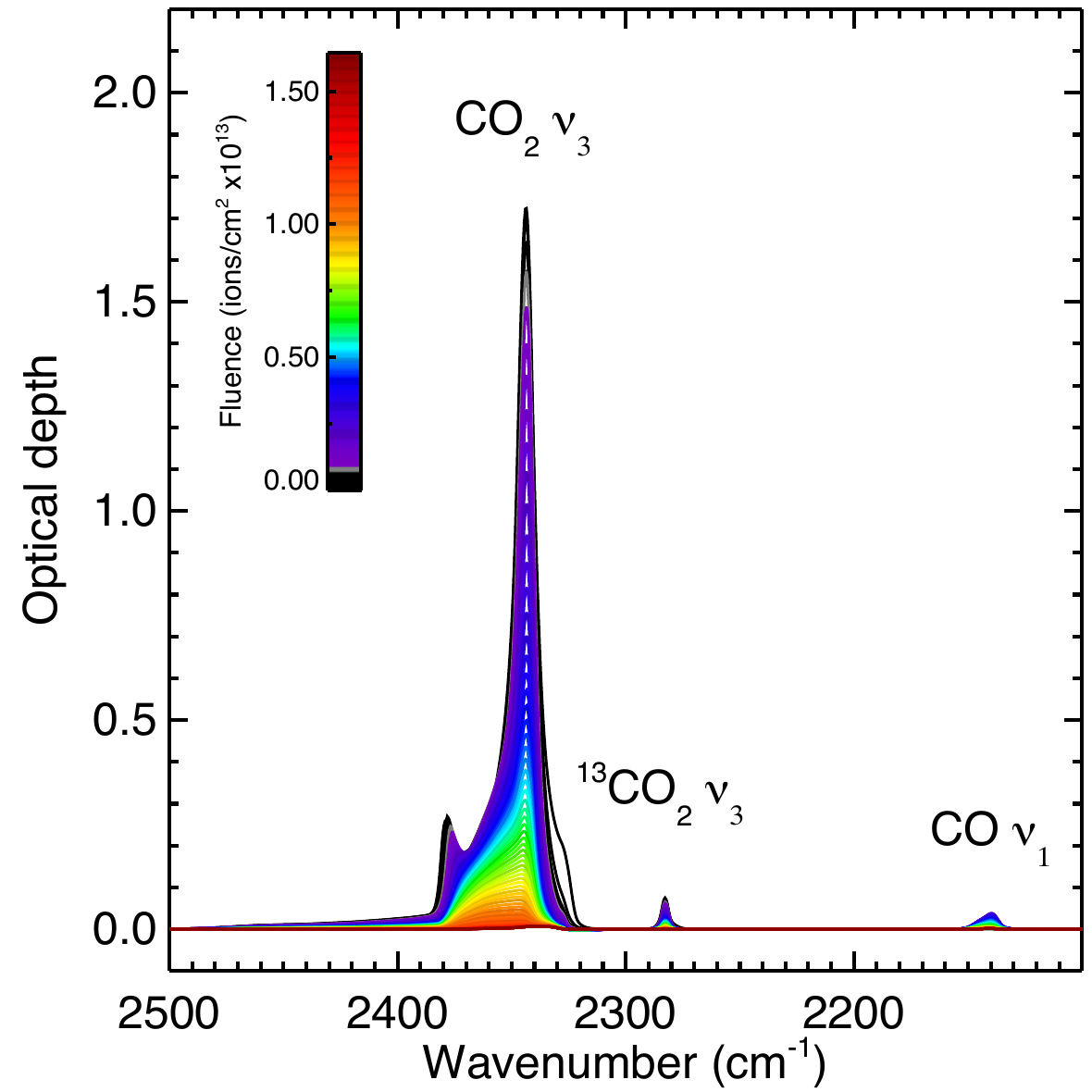}
\caption{Infrared spectra for third CO$_2$ ice experiment with $\rm^{58}Ni^{9+}$ at 33 MeV ions.}
\label{fig:19_05_02_B_W1_BIS_CO2_58_Ni_9plus_0_57_MeV_u}
\end{figure}
%
%
\begin{figure}
\centering
\includegraphics[width=\linewidth]{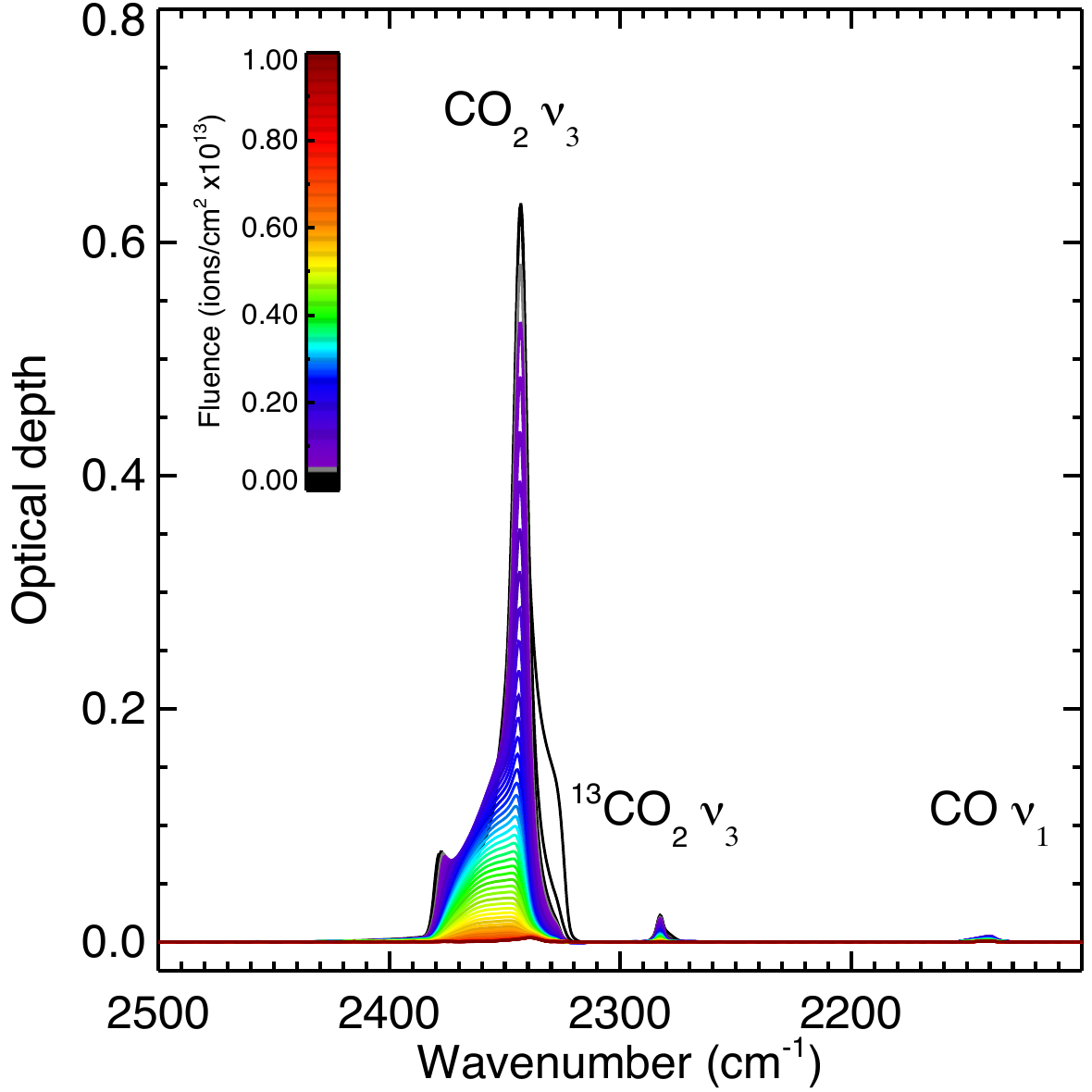}
\caption{Infrared spectra for fourth CO$_2$ ice experiment with $\rm^{58}Ni^{9+}$ at 33 MeV ions.}
\label{fig:19_05_04_D_W1_QUATER_CO2_58_Ni_9plus_0_57_MeV_u}
\end{figure}
%
%
%
\begin{figure}
\centering
\includegraphics[width=\linewidth]{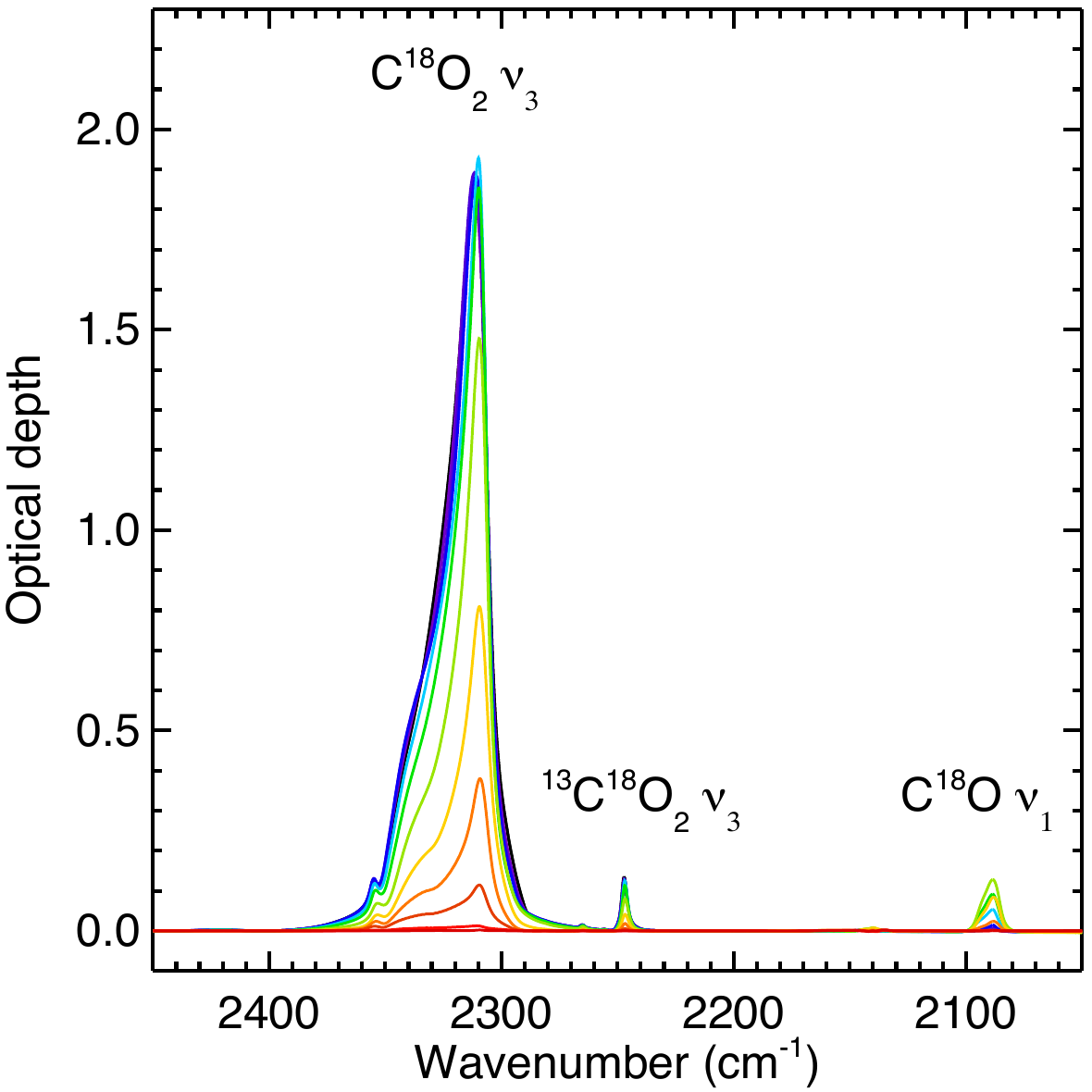}
\caption{Infrared spectra for C$^{18}$O$_2$ ice experiments with $\rm^{58}Ni^{11+}$ at 46 MeV ions.}
\label{fig:C18O2_Ni_46_MeV}
\end{figure}
%
%
%
\begin{figure}
\centering
\includegraphics[width=\linewidth]{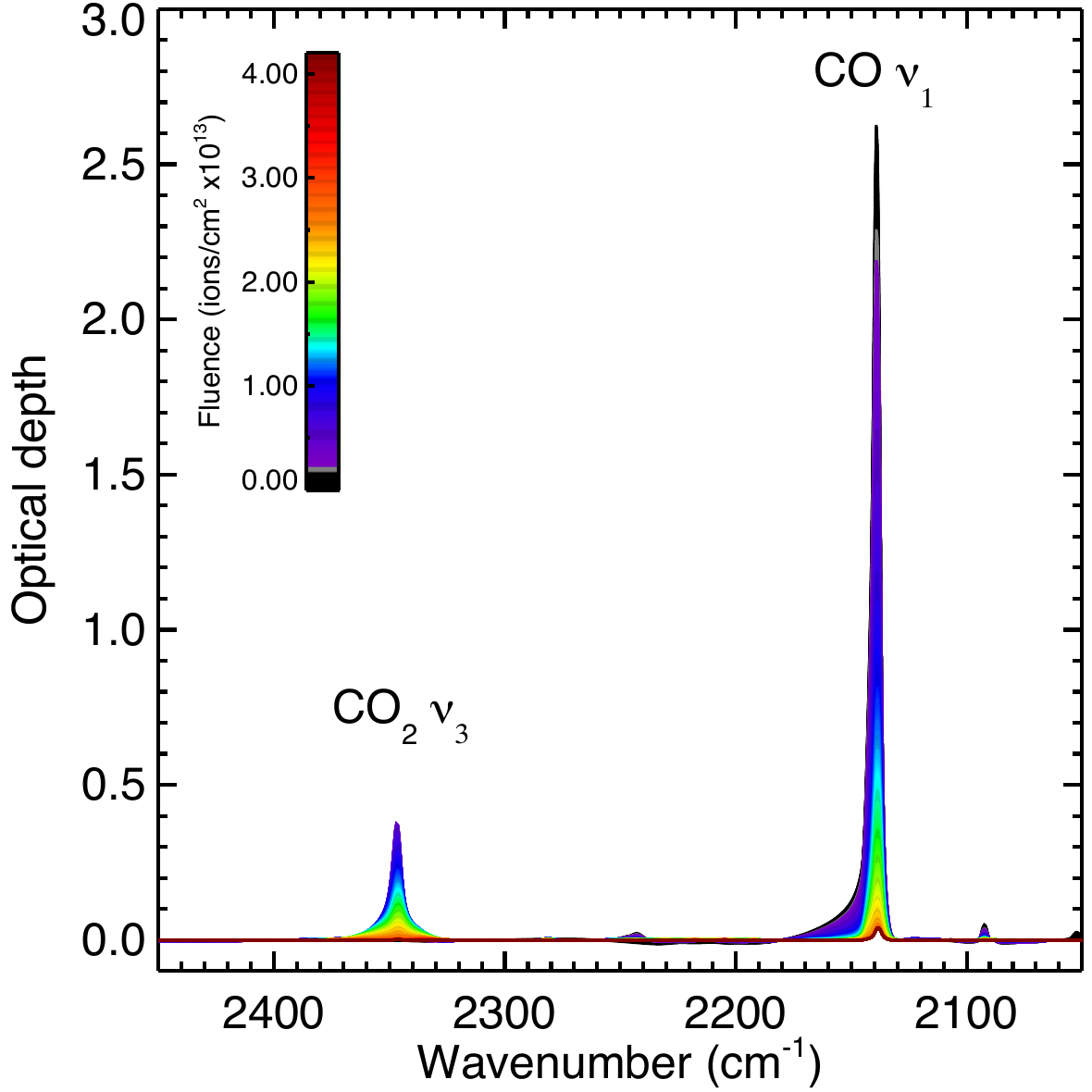}
\caption{Infrared spectra for first CO ice experiment with $\rm^{40}Ca^{9+}$  at 38.4 MeV ions.}
\label{fig:180503_W2_QUATER_CO_40_Ca_9plus_0_96_MeV_u}
\end{figure}
%
%
\begin{figure}
\centering
\includegraphics[width=\linewidth]{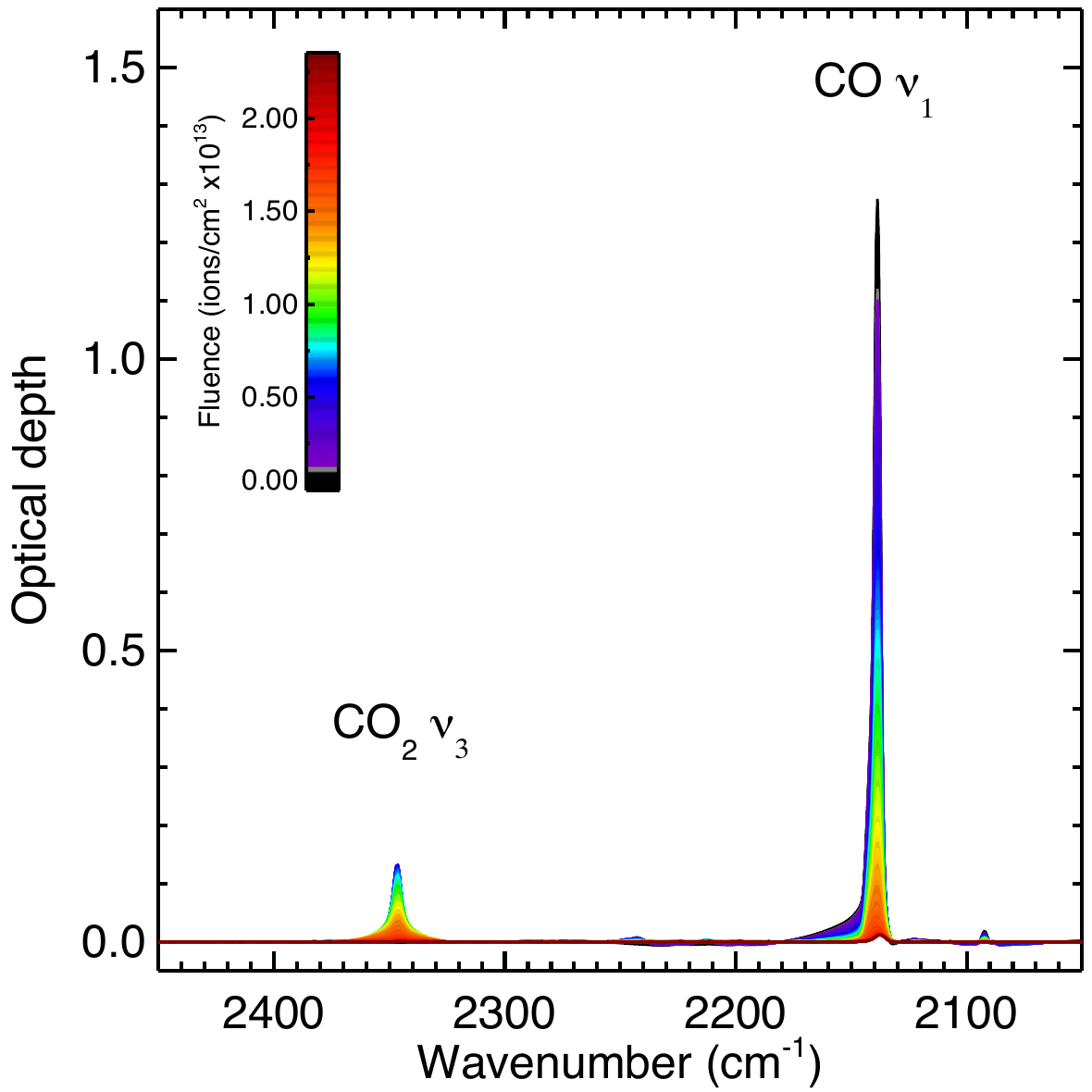}
\caption{Infrared spectra for second CO ice experiment with $\rm^{40}Ca^{9+}$  at 38.4 MeV ions.}
\label{fig:180503_W2_BIS_CO_40_Ca_9plus_0_96_MeV_u}
\end{figure}
%
%
\begin{figure}
\centering
\includegraphics[width=\linewidth]{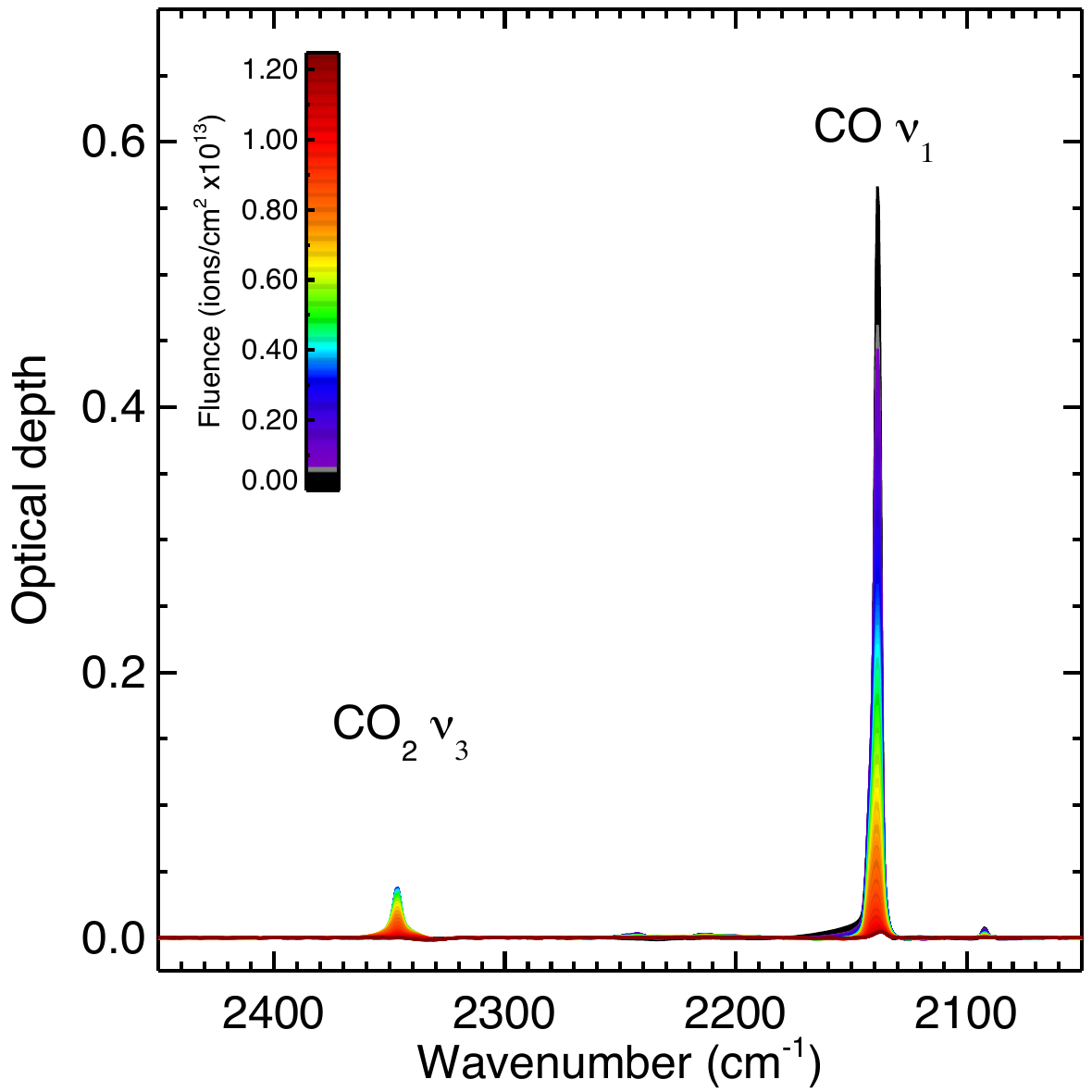}
\caption{Infrared spectra for third CO ice experiment with $\rm^{40}Ca^{9+}$  at 38.4 MeV ions.}
\label{fig:180504_W2_CO_40_Ca_9plus_0_96_MeV_u_spectres}
\end{figure}
%
%
\begin{figure}
\centering
\includegraphics[width=\linewidth]{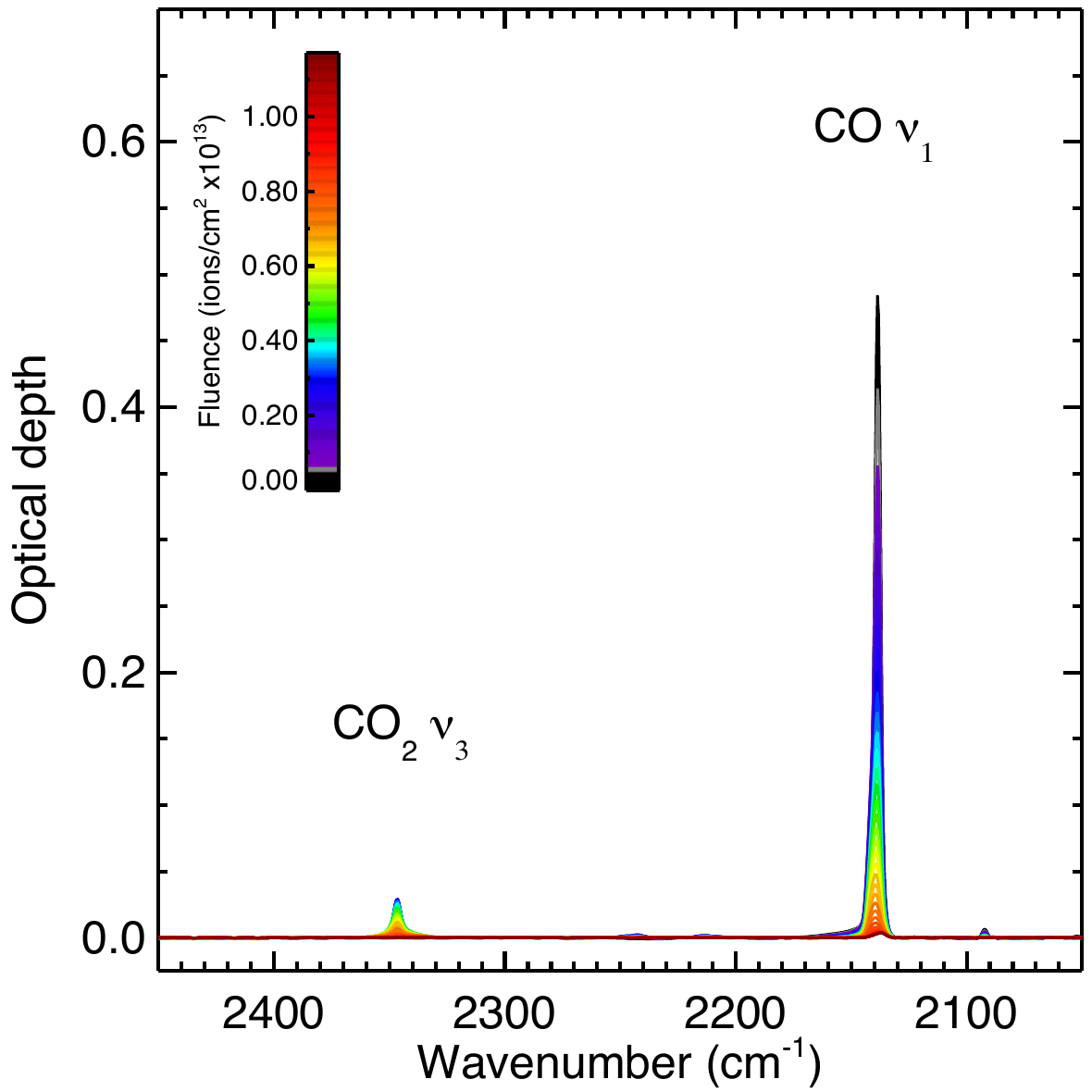}
\caption{Infrared spectra for fourth CO ice experiment with $\rm^{40}Ca^{9+}$  at 38.4 MeV ions.}
\label{fig:180503_W2_TER_CO_40_Ca_9plus_0_96_MeV_u}
\end{figure}
%
%
%
\begin{figure}
\centering
\includegraphics[width=\linewidth]{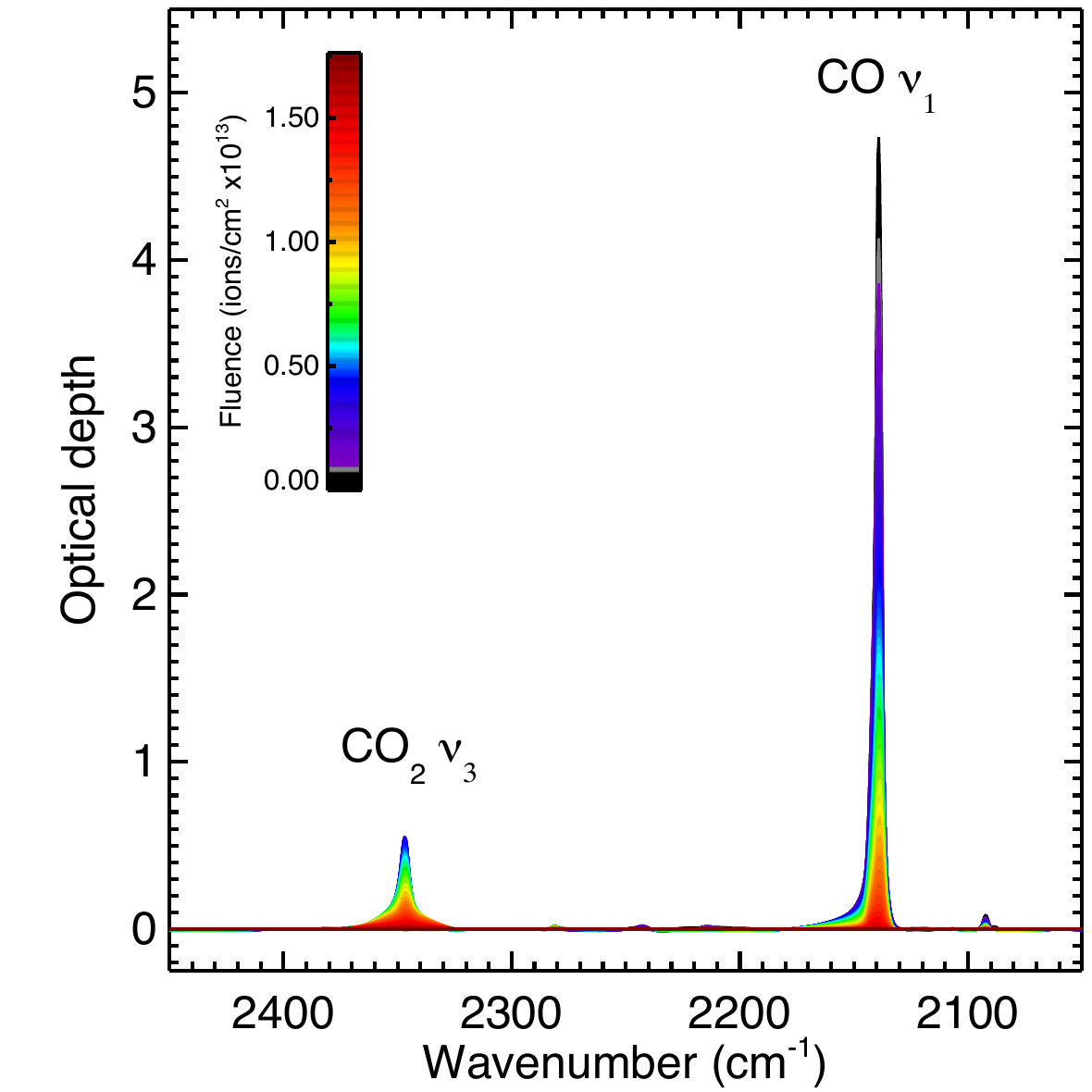}
\caption{Infrared spectra for first CO ice experiment with $\rm^{58}Ni^{9+}$ at 33 MeV ions. We note that the spectra are saturated at the beginning of the experiment. These spectra were discarded from the analysis.}
\label{fig:19_05_04_A_W1_CO_58_Ni_9plus_0_57_MeV_u}
\end{figure}
%
%
\begin{figure}
\centering
\includegraphics[width=\linewidth]{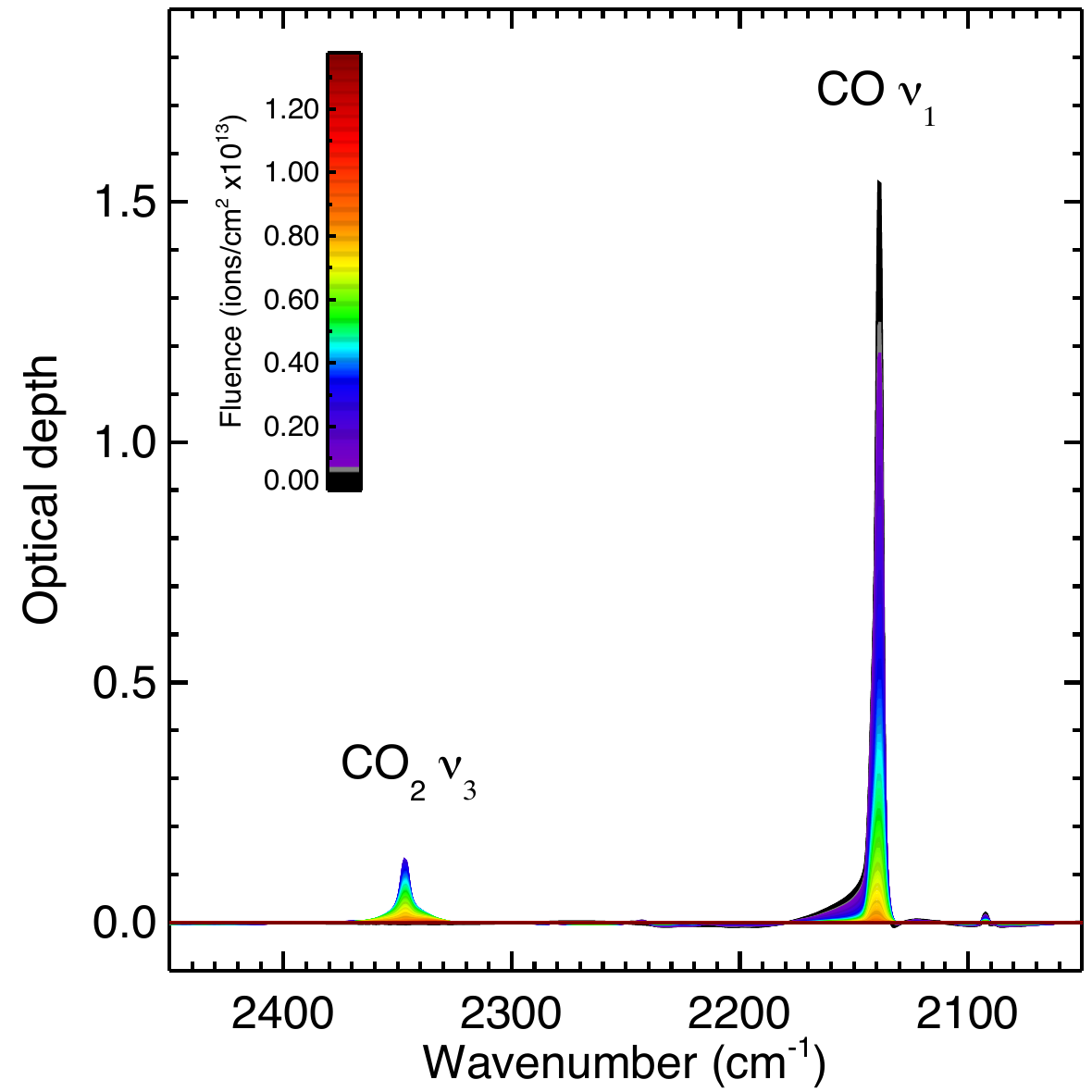}
\caption{Infrared spectra for second CO ice experiment with $\rm^{58}Ni^{9+}$ at 33 MeV ions.}
\label{fig:19_05_04_B_W1_BIS_CO_58_Ni_9plus_0_57_MeV_u}
\end{figure}
%
%
%
\end{appendix}

\end{document}